\documentclass[a4paper,fleqn,usenatbib]{mnras}
\usepackage[T1]{fontenc}
\usepackage{ae,aecompl}

\usepackage{graphicx}	
\usepackage{amsmath}	
\usepackage{amssymb}	

\newcommand{\gadget}{\textsc{Gadget2}}
\newcommand{\popone}{Pop~I}
\newcommand{\poptwo}{Pop~II}
\newcommand{\popthree}{Pop~III}
\newcommand{\hyplot}{\textsc{Hyplot}}

\newcommand{\smalltext}[1]{{\mbox{\tiny #1}}}

\newcommand{\figureref}[1]{Fig.~\ref{#1}}
\newcommand{\figurerefp}[1]{(Fig.~\ref{#1})}

\newcommand{\tableref}[1]{Table~\ref{#1}}
\newcommand{\tablerefp}[1]{(Table~\ref{#1})}

\newcommand{\sectionref}[1]{Section \ref{#1}}

\newcommand{\msol}{~\text{M}_\odot}
\newcommand{\kms}{~\text{km}~\text{s}^{-1}}
\newcommand{\magarcsec}{~\text{mag}~\text{arcsec}^{-2}}
\newcommand{\erg}{~\text{erg}}
\newcommand{\funit}{\times 10^{51}~\text{erg}~\text{M}_\odot^{-1}}
\newcommand{\funitsmall}{\times 10^{50}~\text{erg}~\text{M}_\odot^{-1}}

\title[Constraining subgrid physics in dwarf galaxy simulations]{Constraining the subgrid physics in simulations of isolated dwarf galaxies}

\author[B. Vandenbroucke et al.]{
Bert Vandenbroucke,$^{1}$\thanks{E-mail: bert.vandenbroucke@ugent.be}
Robbert Verbeke,$^{1}$
Sven De Rijcke$^{1}$
\\
$^1$Dept. Physics \& Astronomy, Ghent University, Krijgslaan 281, S9, 9000 Gent, Belgium
}

\date{Accepted XXX. Received YYY; in original form ZZZ}

\pubyear{2015}

\begin{document}
\label{firstpage}
\pagerange{\pageref{firstpage}--\pageref{lastpage}}
\maketitle

\begin{abstract}
Simulating dwarf galaxy halos in a reionizing Universe puts severe constraints on the sub-grid model employed in the simulations. Using the same sub-grid model that works for simulations without a UV-background (UVB) results in gas poor galaxies that stop forming stars very early on, except for halos with high masses. This is in strong disagreement with observed galaxies, which are gas rich and star forming down to a much lower mass range. To resolve this discrepancy, we ran a large suite of isolated dwarf galaxy simulations to explore a wide variety of sub-grid models and parameters, including timing and strength of the UVB, strength of the stellar feedback, and metallicity dependent Pop III feedback. We compared these simulations to observed dwarf galaxies by means of the baryonic Tully-Fisher relation (BTFR), which links the baryonic content of a galaxy to the observationally determined strength of its gravitational potential. We found that the results are robust to changes in the UVB. The strength of the stellar feedback shifts the results on the BTFR, but does not help to form gas rich galaxies at late redshifts. Only by including Pop III feedback are we able to produce galaxies that lie on the observational BTFR and that have neutral gas and ongoing star formation at redshift zero.
\end{abstract}


\section{Introduction}
Dwarf galaxies are the lowest-mass inhabitants of the extragalactic Universe. Their haloes are predominantly made up of dark matter (DM), which makes them the ideal probing ground for dark matter or alternative gravity theories. Moreover, their relatively shallow gravitational potential makes their evolution very sensitive to all sorts of internal and external physical processes, ranging from stellar feedback by supernovae to the heating by an external UV-background (UVB). For this reasons, numerical simulations of the formation and evolution of dwarf galaxies provide a promising way to solve many questions about the existence and character of dark matter \citep{oh, vogelsberger_dwarfs}, the strength and timing of the UVB \citep{simpson, sawala} and the specifics of star formation and stellar feedback \citep{annelies_sfp, joeri, fire}.

There are two major classes of numerical simulations of dwarf galaxies. Cosmological zoom-simulations start with a large cosmological box filled with dark matter and optionally baryons, and locally increase the resolution by means of refinement levels \citep{shen, vogelsberger_dwarfs, sawala}. Some of these simulations focus on more massive haloes and mainly study dwarf galaxies as satellites or close companions of these haloes. These satellite haloes are influenced by the presence of their host and it remains unclear if this makes them fundamentally different from the isolated field dwarf galaxies \citep{sawala}. Since the best observational data about dwarf galaxies comes from Milky Way satellites, this type of constrained simulations is very useful. There are also zoom simulations of individual isolated dwarf galaxies \citep{vogelsberger_dwarfs} or groups of dwarf galaxies \citep{shen}. Setting up a cosmological zoom simulation however requires a careful selection of an appropriate zoom region, which is an expensive process that requires a suite of preliminary simulations to be run and even then allows for little control over the final properties of the selected halo.

Simulations of isolated dwarf galaxy haloes either start with an idealized set-up containing a dark matter halo with a theoretically derived density profile and a gas halo embedded therein to study the self-consistent formation of the galaxy \citep{stinson, annelies_sfp, revaz, joeri, robbert}, or start with a fully self-consistent galaxy including a stellar disk and subject this galaxy to external processes \citep*{lokas}. Because these simulations only contain the mass that resides in the dwarf halo, it is possible to obtain much higher resolutions than for zoom-simulations, with less computational cost. This also allows for a large parameter survey where not only model parameters but also structural parameters can be easily changed. However, these simulations do not take into account cosmological effects like the large scale gravitational potential or the effects of mass accretion, apart from the initial density profile, which is derived from cosmological simulations. This drawback can be alleviated by running merger simulations which take into account the merger history of the isolated halo \citep{annelies_mergers}.

The ultimate goal of numerical dwarf galaxy simulations is to produce model data that can be compared to observed dwarf galaxies to constrain theoretical models. We therefore need to find model quantities that can be easily and unambiguously observed. An often quoted relation is the so called $M_\smalltext{star}-M_\smalltext{halo}$ relation \citep*{guo, behroozi,magicc,moster}, which links haloes from large scale cosmological (zoom-)simulations to observed stellar mass aggregates by means of an abundance matching technique. This relation is not applicable in the dwarf regime however, since (a) halo masses cannot easily nor unambiguously be observed, and (b) the starting premise of the technique, i.e. the most massive halo corresponds to the most luminous stellar aggregate, breaks down at low masses \citep{sawala}. A more accessible tracer of the halo potential is the circular velocity, which can be measured from resolved rotation curves. Together with the observed stellar and neutral gas mass, this quantity yields the so-called baryonic Tully-Fisher relation (BTFR) \citep{mcgaugh}. This purely observational relation is a more promising candidate for model comparisons, since it relates a halo property to baryonic properties. \cite{brooks_zolotov} discuss the position of a set of simulated dSph satellites on the Tully-Fisher relation, but to our knowledge, no simulated BTFR in the dwarf regime has been published before.

Without an UVB, numerically producing dwarf galaxies that have realistic observational properties poses no real challenge \citep{sander}, especially since there is a degeneracy between the stellar feedback parameters and the position of the resulting dwarf galaxy on a range of observational scaling relations \citep{annelies_sfp}. However, these systems tend to have unrealistically high stellar masses and bursty SFHs \citep{joeri}. When the UVB is taken into account, the evolution of these galaxies looks a lot different~: star formation is limited to very early in the simulation, after which the galaxies become gas poor and star formation stops \citep{simpson}. Or star formation is delayed until very late in the simulation, yielding very exotic dwarf galaxies \citep{shen}. Cosmological zoom-simulations predict that only half of the dark matter halos with circular velocities of $\sim 25 \kms$ host galaxies that are able to form stars, let alone keep neutral gas, and this fraction rapidly decreases towards lower circular velocities \citep{sawala}. This is in disagreement with observed dwarf irregular galaxies (dIrrs), which have continuous star formation histories \citep{lcid_monelliA, weiszA}, are able to keep neutral gas throughout their evolution, and show no clear imprint of reionization \citep{lcid_monelliB, weiszB}. The number density of these systems roughly equals that of the dark matter halos from cosmological simulations \citep{tollerud}, indicating that much more dark matter halos host gas rich, star-forming, low mass dwarf galaxies than predicted. Furthermore, ultra-faint dIrrs like Leo~P \citep{giovanelli}, Leo~T \citep{irwin} and Pisces~A \citep{tollerud} have a large neutral gas content and circular velocities of $\sim15\kms$.

This disagreement has two possible explanations~: (a) there is something wrong with the simulations and especially the sub-grid model that is employed, or (b) there is something wrong with how models and observations are compared, e.g. the halo masses inferred from observations are lower than the corresponding halo masses in simulations.

In this work, we will run a large suite of simulations of isolated dwarf galaxy haloes to try to reproduce the observed BTFR and address this problem. We will discuss the differences between theoretical quantities and the quantities that are actually observed and try to produce mock observations that resemble the real observations as closely as possible. This should allow us to reliably compare our models with observed field dIrrs and allow us to constrain the details of our galaxy evolution model. We will focus on three different aspects of our model~: the stellar feedback strength, the strength and timing of the external UVB, and the specifics of a new population III feedback model. We will explain how these parameters affect the galaxy star formation history (SFH) and how this influences the final position of the galaxy on the BTFR.

The paper is structured as follows~: \sectionref{section_model} gives the details of our simulations and the model parameters. \sectionref{section_analysis} gives details about our analysis of the results. \sectionref{section_results} gives an overview of the results for the different parameters, including a discussion of the effect of stochasticity and resolution on the final result. Finally, \sectionref{section_conclusion} will present our conclusions.

\section{Model}\label{section_model}
We ran a large suite of numerical simulations of isolated dwarf galaxy halos using an adapted version of the N-body/SPH code \gadget{} \citep{gadget_paper}. The adaptations consist of a model for star formation using sink particles \citep{sander}, gas cooling and heating using a 5 parameter model that includes metal dependent cooling and heating by an ionizing UVB \citep{sven_cooling_curves}, stellar feedback from SNII, SNIa, stellar winds and optionally population III stars, and an advanced equation of state that takes into account the ionization state of the gas \citep{mezelf}.

In this section, we describe the initial conditions of our models, and discuss the different subgrid models and parameters that are varied during our study. We conclude with a general overview of our models and simulations.

\subsection{Initial conditions}

\subsubsection{Isolated set ups}

The initial conditions for our isolated halos are generated using a Monte Carlo sampling technique for both the initial DM-halo and gas halo. The former is set up as a NFW-halo \citep*{nfw} with a concentration parameter given by \citep{annelies_mergers}~:
\begin{equation}
c \approx 33 \left( \frac{M_\smalltext{h}}{10^8\msol{}} \right)^{-0.06},
\end{equation}
with $M_\smalltext{h}$ the total mass of the halo.
The latter is set to be in pseudo thermal equilibrium, with a density profile of the form \citep{joeri}~:
\begin{equation}
\rho_\smalltext{gas} (r) = \frac{\rho_{\smalltext{gas},c}}{1+\left(\frac{r}{r_c}\right)^2},
\end{equation}
where $r_c$ is set to equal the scale length of the NFW-halo, while the central density $\rho_{\smalltext{gas},c}$ is related to the central density of the NFW-halo through
\begin{equation}
\rho_{\smalltext{gas},c} = \frac{\Omega_\smalltext{b}}{\Omega_\smalltext{DM}} \rho_{\smalltext{DM},c},
\end{equation}
with $\frac{\Omega_\smalltext{b}}{\Omega_\smalltext{DM}}=0.2115$ \citep{WMAP}.

Initially, the particles that sample the dark matter receive random velocities drawn from the isotropic distribution function corresponding to the NFW density profile. To prevent these velocities from erasing the NFW cusp in the scarcely sampled central part of the distribution, we construct so-called `quiet' initial conditions, in which the velocities are assigned in a symmetric way \citep{annelies_sfp}.

The gas halo is initially in rest, but is optionally set up with a constant solid body rotation, to study the effect of angular momentum on the dwarf galaxy \citep{2013Schroyen}. The rotation velocity $v_\text{rot}$ is a model parameter.

All models are run with 50,000 dark matter particles and 50,000 smoothed particle hydrodynamics (SPH) gas particles (which we will call \emph{low resolution} runs). To verify that this is indeed enough to obtain a qualitatively correct behaviour of the resulting dwarf galaxy, we ran some of our models with 200,000 particles of each type (\emph{high resolution} runs). Note that low and high resolution do not explicitly denote an actual physical resolution difference, but rather a fixed relative resolution difference of a factor of 4. The actual mass resolution for the particles is set by the desired total mass of the dark matter halo. The smoothing length of the gas particles is dynamically set by requiring a fixed number of $50\pm 1$ neighbours for all particles. The softening length of all components (DM, gas and stars) is set to a fixed value, which roughly corresponds to the smoothing length of a gas particle that has a density equal to the density treshold for star formation (calculated as the radius of a sphere with this density, containing 50 gas particles). For clarity, the mass resolution and softening lengths for the different models have been specified in \tableref{table_ics}.

Our models start at a redshift of 12, which corresponds to a lookback-time of 13.37 Gyr with the cosmological parameters employed \citep{WMAP}, and are run until a redshift of 0, or until the total computation time exceeded the arbitrary limit of 3 months, which happened for some of the high resolution simulations due to physical reasons. All simulations were run on our local computing infrastructure, which consists of 5 computing nodes with Intel CPUs with between 16 and 64 physical cores.

The initial conditions for our simulations contain only 3 free parameters~: the total halo mass $M_h$ that is sampled, the initial rotation velocity $v_\smalltext{rot}$ and the number of particles used to sample the distributions.

\subsubsection{Merger simulations}

Due to the idealized set up of the isolated initial conditions, we might potentially miss two important effects that might shape the BTFR~: the accretion of cold gas that will fuel star formation and the angular momentum acquisition of the halo due to mergers. To effectively resolve these effects, cosmological zoom simulations are needed. It is however very computationally expensive to run a parameter study of the size considered in this work with this type of simulations.

To get an idea of the effect of including these effects, we will also consider a merger simulation. Instead of simulating a single isolated halo, we set up a system of smaller halos, that are then allowed to merge according to a merger tree that is sampled using the extended Press-Schechter theory, with a conditional mass function that is fitted to the merger trees from the Millenium simulation \citep*{annelies_mergers, verbeke2015}. This allows us to incorporate cosmological effects into our simulations without using full blown cosmological zoom simulations. The halos we consider are relatively light, so we expect them to only accrete gas through this type of mergers with other small halos, since they lack the mass to acquire unbound UVB heated gas.

\subsection{Sub grid models}

\subsubsection{UV background}\label{subsection_UVB_model}
The standard UVB we use is the one provided by \citet{faucher_giguere}, which kicks in at a redshift of 10.5  and reaches its peak strength at a redshift of 2. Since our runs start at a redshift of 12, the start of the UVB coincides with the onset of star formation in our models. We therefore also experimented with a UVB that only starts at a redshift of 7 (we will call this \emph{late UVB} runs, contrary to the standard \emph{early} UVB). To qualify the effect of the UVB on the star formation, we also ran models with a UVB with a strength that is only 10 per cent of the normal strength (\emph{low UVB} runs).

The UVB acts as a heating term \citep{sven_cooling_curves} and also influences the ionization equilibrium, which introduces a redshift dependence for the cooling curves and the multiphase equation of state (this is an extension of the model of \cite{mezelf}). The latter takes into account the potential energy reservoir connected to the ionization of (mainly) hydrogen, which breaks the linear dependence of the gas temperature on the internal energy of the gas. This takes care of the thermal energy absorbed by the ionization of neutral gas in a subgrid fashion, but does not significantly affect our models.

\subsubsection{Star formation rate and stellar feedback}\label{subsection_model_stellar_feedback}
Star formation is modelled using the approach described by \cite{sander}. If a gas particle is in a region of converging gas flow, has a temperature below 15,000~K and a density above a density threshold of 100 amu cm$^{-3}$ \citep{joeri}, it is potentially converted into a star particle, which is dynamically equal to a DM particle, but has some extra properties attached to it (metallicity, formation time,...). These star formation criteria together with the multiphase equation of state guarantee that only cold, neutral and collapsing gas can form stars. The probability with which this conversion happens is stochastically sampled to reproduce a Schmidt law of the form
\begin{equation}
\frac{\text{d}\rho{}_\text{stars}}{\text{d}t} = c_* \frac{\rho{}_\text{gas}}{t_\text{dyn}},
\end{equation}
where $\rho{}_\text{stars}$ and $\rho{}_\text{gas}$ are the density of the stars and the gas, and the dynamical time $t_\text{dyn}$ is given by
\begin{equation}
t_\text{dyn} = \frac{1}{\sqrt{4\pi{}G\rho{}_\text{gas}}}.
\end{equation}

The star formation efficiency is set by a single parameter, $c_*$. However, this parameter is degenerate due to the self-regulating character of star formation \citep{stinson_2006}. A lower star formation efficiency leads to a larger initial star formation peak, since gas is converted into stars more slowly and more gas reaches the density threshold before stellar feedback shuts down further star formation efficiently. This in turn leads to a higher stellar feedback at later times, which suppresses later star formation. The star formation efficiency parameter is therefore also coupled to another parameter in our system~: the stellar feedback efficiency, which sets the fraction of the stellar feedback that is effectively absorbed by the interstellar medium. This parameter is closely related to the numerical restrictions of our feedback model~: since we do not resolve the hot gas bubbles around supernova explosions, we do not heat the gas particles in our simulation efficiently enough and most of the energy we put in is radiated away by our cooling model \citep{dalla_vecchia}. Because of this, it is unclear how much energy ultimately ends up in the interstellar medium. Hence the feedback efficiency parameter. To prevent severe over-cooling, we shut off radiative cooling for all gas particles that receive direct stellar feedback from SNII and SW, allowing the feedback to spread out adiabatically. For SNIa feedback we do not shut off radiative cooling, since this feedback typically affects gas that is more hot and diffuse and does not suffer from over-cooling. We also present a model in which we do not switch off cooling to illustrate the effect of over-cooling.

Due to the degeneracy of the star formation efficiency and feedback efficiency parameters, it suffices to only vary one, while keeping the other fixed. We choose to fix $c_*$ at a value of 0.1 \citep{annelies_mergers}.

Since this resolution issue affects all types of feedback, we use the same feedback efficiency parameter to scale all energy injections due to stellar feedback, irrespective of the feedback type. Different types of feedback then still differ in relative strength and timing, which will cause them to affect the simulations in distinct ways.

Star particles in our simulations return three distinct types of feedback to the surrounding gas during their lifetime. Immediately after the star particle has formed, massive stars with masses in the range $8-70\msol{}$ will return high energetic radiation to the surrounding ISM in the form of stellar winds (SW). Following \citet{sander}, we insert $1.0\funitsmall{}$ energy due to SW in the surrounding ISM, weighted with the fraction of the mass of a single stellar population (SSP) that is in the range $8-70\msol{}$ for a Chabrier IMF ($1.18\times 10^{-2}$) \citep{chabrier} and multiplied with the feedback parameter. This energy injection is spread out uniformly over a time interval $0-31$~Myr, corresponding to the lifetime of these massive stars.
When these massive stars reach the end of their lifetime, they explode in SNII explosions, which return an energy of $1.0\funit{}$ to the ISM, again weighted with the SSP fraction and the feedback parameter. This happens in a time interval $3.8-31$~Myr and is also done uniform in time.

Next to energy, SNII also return material to the ISM. Some of it is in the form of H and He, but part of it is also in the form of heavy metals, that enrich the ISM. For every SSP, we return a fraction of 0.191 of its mass to the surrounding ISM \citep{sander}. For every gas particle, we also keep track of the mass in Fe and Mg (the two metallicity tracers used for the cooling and heating and the advanced gas physics). For every SSP, we return a fraction $9.33\times 10^{-4}$ of its mass as Fe and a fraction of $1.51\times 10^{-3}$ of its mass as Mg. All matter is returned uniformly during the same time interval as the SNII energy.

Stars with masses less than $8\msol{}$ do not give rise to SNII explosions. However, less massive stars can form white dwarfs and if they are part of a binary system, mass overflow between a red giant and its companion white dwarf can cause the latter to exceed its Chandrasekhar limit, giving rise to a SNIa explosion. If we assume that all stars in the mass range $3-8\msol{}$ form white dwarfs, we can estimate the number of possible SNIa explosions from the IMF. This then has to be multiplied with an extra factor to take into account that not all white dwarfs will be part of a binary system and not all binary systems will experience the necessary mass overflow. We use a fixed ratio of SNIa to SNII explosions of 0.15 \citep{sander}. These SNIa explosions also release an energy of $\sim 1.0\funit$ to the surrounding ISM.

Due to the large variety of companion masses in binary systems, SNIa feedback is spread out over a large interval in time. We use the Gaussian model of \citet{strolger} and return the total energy using a normal distribution centered on a delay time of $\tau=4$~Gyr and with a standard deviation of $\frac{1}{5}\tau$ \citep{napoleon}. To limit the computational overhead caused by returning very low feedback values in the tails of the normal distribution, we limit feedback to a $3\sigma$ time interval around the Gaussian peak value.

SNIa also return mass and metals to the ISM. We return a fraction of $6.55 \times 10^{-3}$ of the mass of the SSP in mass, a fraction of $1.65\times 10^{-3}$ of its mass as Fe, and a fraction of $2.58\times 10^{-4}$ as Mg \citep{sander}.

When energy or mass is transferred between a star particle and the surrounding ISM, we will always spread it out across 50 neighbouring SPH particles of the star particles. To this end, we associate a smoothing length to every star particle and iteratively look for neighbouring SPH particles until 50 neighbours are found within this smoothing length. The energy or mass is then spread out according to the same kernel function that is also used for the SPH density calculation and using this smoothing length. To ease the iteration, a small deviation in the number of neighbours is allowed. We set this deviation to 1 for all our runs.

\subsubsection{\popthree{} feedback}\label{subsection_popthree_models}
In all our runs, the gas starts of with zero metallicity and hence we can consider star particles formed out of this zero metallicity gas to be very metal poor ``population three'' or \popthree{} stars. We can also relax this characterization to include stellar particles with very low, non zero metallicities (we use the arbitrary upper limit $[\text{Fe}/\text{H}] < -5$).

\popthree{} stars are still poorly constrained, mainly because there are no observational data and all our knowledge has to come from numerical simulations \citep{susa,nomoto}. Although most models do suggest that they can obtain very high masses, there is no consensus on the form of the \popthree{} initial mass function (IMF), nor on the energy output of \popthree{} supernova (SN) explosions. \citet{nomoto} composed tables of energy and metal yields for massive stars, including \popthree{} stars, up to $300\msol{}$. These energies are very different from the typical energies released by normal ``population II'' (\poptwo{}) and ``population I'' (\popone{}) supernovae, and hence will probably affect the ISM in the simulations in a different way.

\paragraph*{Model 1}
As a first method to include \popthree{} stars, we just assumed \popthree{} stars to be star particles with zero metallicity and with masses in the range $60-300\msol{}$. The lower limit was chosen arbitrarily and is close to the upper limit we use for the masses of \poptwo{} and \popone{} stars. As a first approximation, we only look at the energy output of the \popthree{} supernovae and not their metal output. To this end, we scale up the number of supernova explosions per star particle (which represents a stellar population) to match the total energy output of a \popthree{} stellar population with the same total mass, but we keep the energy and metal output per supernova fixed.

Even in this simple model, there are already some uncertainties coming from the uncertainty on the \popthree{} IMF. We can assume the same Chabrier IMF we use for the \poptwo{} and \popone{} stars for the \popthree{} stellar population and extrapolate it out to the upper mass limit $300\msol{}$, but we can also use a flat IMF in the mass range $60-300\msol{}$, which seems to be in better agreement with numerically found IMFs for \popthree{} stars \citep{susa}. Since the number of massive stars in the latter case will be a lot higher, this will have a major impact on the total energy output of the \popthree{} stellar population. We will call this model ``\popthree{} model 1A'', and parametrize it by the fraction of the total mass of the star particle that consists of high mass stars that will explode as supernovae and each return $10^{51}\erg{}$ of energy to the ISM. The total energy a star particle pumps into the ISM (over the whole feedback time interval) is then given by multiplying this parameter with the mass of the stellar particle. We will consider a model with fraction $0.06358\funit{}$ (low feedback), $4.9025\funit{}$ (high feedback), and $0.1467\funit{}$ (middle feedback). The high feedback value illustrates the fact that for a flat IMF, the total energy emitted by a \popthree{} SN is a lot higher than for a SNII. The time interval for this feedback corresponds to the expected lifetime of the lower and upper mass limit stars~: 0.006 to 0.36 Myr \citep{nomoto}.

We also investigate the effect of the lower mass limit for \popthree{} stars on the \popthree{} energy feedback, by considering a model with lower limit $140\msol{}$ instead of $60\msol{}$. In this case, all \popthree{} stars will be extremely short-lived, and will deposit a large amount of energy into the ISM over a very short time interval (0.006 to 0.043 Myr). We will call this model ``\popthree{} model 1B'' and consider two feedback parameters~: $0.1814\funit{}$ (low feedback), and $0.32\funit{}$ (high feedback).

\paragraph*{Model 2}
A second, more advanced \popthree{} model also takes into account the energy radiated away by the \popthree{} stars during their stellar lifetime, which is similar to the energy output provided by SW for \poptwo{} and \popone{} stars, but is potentially a lot higher \citep{heger_and_woosley}. This energy output bridges the very short gap between the birth of the star particle and the onset of \popthree{} SN feedback and provides a further early energy input into the ISM. We consider two versions of this model~: a model with high SW energy ($10^{52}\erg{}$) and low \popthree{} SN feedback ($0.007361\funit{}$), and a model with low SW energy ($10^{51}\erg{}$) and high \popthree{} SN feedback ($0.051765\funit{}$).

\paragraph*{Model 3}
A third and most advanced \popthree{} model also takes into account the metal yields provided by \citet{nomoto} to provide a more realistic metal enrichment of the ISM by \popthree{} supernovae. This model combines the knowledge obtained from the previous two models with a fully consistent treatment of \popthree{} stars, and is used by \citet{verbeke2015}. We will verify this model as part of our parameter study. Apart from returning the low \popthree{} SW feedback and high \popthree{} SN feedback to the ISM, a \popthree{} star particle returns a fraction of 0.45 of its total mass to the surrounding ISM. Most of this mass is in the form of H and He, but a fraction of 0.026 is in the form of metals, with a total Fe mass fraction of $9.327\times 10^{-5}$ and a Mg mass fraction of $1.514\times 10^{-4}$. These last values are simply 10 per cent of the corresponding value for a normal SNII, and were based on the metals yields of \citet{nomoto}.

\subsection{Simulation overview}
In total, 263 simulations were run, with a total of 15 different combinations of code versions and parameter values as discussed above. Apart from this, we considered initial conditions with 5 different halo masses ($1\times 10^9 \msol{}$, $3\times 10^9 \msol{}$, $5\times 10^9 \msol{}$, $7\times 10^9 \msol{}$ and $9\times 10^9 \msol{}$), 3 halo rotation velocities (no rotation, $5 \kms{}$ and $10 \kms{}$) and 2 resolutions ($2\times50,000$ and $2\times 200,000$ particles). We adopt a simple naming convention for our simulations, based on (a) the code and parameters with which the simulations was performed \tablerefp{table_codes}, and (b) the parameters of the initial conditions \tablerefp{table_ics}. The name for a simulation of a $1\times 10^9 \msol{}$ model with a low resolution and no rotation with a code with no \popthree{} stars and a late and low UVB, and with a low stellar feedback parameter e.g. will be C1P1M1R00L.

We only have 30 different initial conditions (ICs), of which 15 low resolution ICs that were used for almost all models. To check the effect of stochastic changes to the simulations due to Poisson noise in these ICs, we reran 1 model with ICs that were generated using a different random seed. This model is called C1P1bis and has code and parameter values equal to those of C1P1.

For the models C1P1 and C3P2, we ran all simulations using both the low resolution and the high resolution initial conditions, to check the convergence of our simulations. For the other models, we only ran one high resolution simulation, M9R10H, and use this as a general check on the convergence of that specific model.

Apart from the initial conditions described above, we also discuss one merger simulation, to assess the influence of cosmological effects on our models. This merger simulation is part of a suite of merger simulations that was run as a result of the parameter study performed in this work \citep{verbeke2015}. For clarity, we will keep its name from that work~: DG10e9-NP3. This model has the same properties as model C3P1~: a full and early UVB, feedback efficiency 0.7 and no \popthree{} feedback. We will denote it by a $+$ symbol.

In the end, 7 out of 263 simulations exceeded the 3 month time limit imposed~: models C1P1M7R00H, C1P1M9R00H and C1P1M9R05H, model C3P2M9R00H, and models CeP1M5R00L, CeP1M7R00L and CeP1M9R10H. All these models form an excessive number of stars from the beginning of the simulations, leading to a very computationally expensive stellar feedback contribution, which explains the long run time. However, this excessive star formation is in line with other results for the same code and parameter values, so we learn nothing new from these simulations. We will discard them anyway.

3 simulations crashed~: CeP1M1R00L, CeP1M3R00L and CeP1M9R00L. These simulations did not apply an adiabatic cooling periodic to gas that received stellar feedback and as a result formed so many stars that the program did no longer find any gas to give feedback to and crashed. Similar behaviour was found for the other simulations of this model that did successfully run, so that we can discard these simulations.

Model C1P1M9R00H shows very strange behaviour, in the sense that while for all other models the circular velocity tends to decrease over time, it increases significantly for this model, ending up with an unrealistically high circular velocity of more than 150$\kms$. The low resolution equivalent C1P1M9R00L has a more realistic circular velocity but ends up with almost no neutral gas, contrary to similar models with the same mass and different rotation parameters or the same rotation parameter and lower masses, which still have neutral gas at the end. Simulation C1P1bisM9R00L, which has the same parameters but a differently sampled IC shows the same behaviour. We conclude that something is fundamentally wrong with model C1P1M9R00 and will therefore discard these 3 simulations. For very similar reasons, we will also discard simulation C3P2M9R00L. Other models with the M9R00L IC will be taken into account.

From the 263 simulations, 250 will hence be discussed in the remainder of this paper.

\begin{table*}
\centering
\caption{Code and parameter values naming convention.\label{table_codes}}
\begin{tabular}{|lclclc|}
\hline
Code & Symbol & UVB model & $f_\text{feedback}$ & \popthree{} model & number of simulations \\
\hline
C1P1 & $\blacktriangle$ & low and late & 0.7 & no \popthree{} stars & 30\\
C1P1bis & \textcolor[gray]{0.5}{$\blacktriangle$} & low and late & 0.7 & no \popthree{} stars & 15\\
C2P1 & $\diamond$ & full and late & 0.7 & no \popthree{} stars & 16\\
C3P1 & $\blacksquare$ & full and early & 0.7 & no \popthree{} stars & 16\\
C3P2 & \textcolor[gray]{0.5}{$\blacksquare$} & full and early & 1.0 & no \popthree{} stars & 30\\
C3P3 & $\square$ & full and early & 2.0 & no \popthree{} stars & 16\\
C4P2 & $\circ$ & no UVB & 1.0 & no \popthree{} stars & 16\\
C7P4 & $\blacktriangleleft$ & full and early & 0.7 & \popthree{} model 1A: low feedback & 16\\
C7P6 & \textcolor[gray]{0.5}{$\blacktriangleleft$} & full and early & 0.7 & \popthree{} model 1A: high feedback & 16\\
C7P7 & $\triangleleft$ & full and early & 0.7 & \popthree{} model 1A: middle feedback & 16\\
C9P8 & $\blacktriangleright$ & full and early & 0.7 & \popthree{} model 1B: low feedback & 16\\
C9P9 & \textcolor[gray]{0.5}{$\blacktriangleright$} & full and early & 0.7 & \popthree{} model 1B: high feedback & 6\\
CaPa & $\blacktriangledown$ & full and early & 0.7 & \popthree{} model 2: high stellar winds, low feedback & 16\\
CbPc & $\spadesuit$ & full and early & 0.7 & \popthree{} model 2: low stellar winds, high feedback & 16\\
CcPd & $\clubsuit$ & full and early & 0.7 & \popthree{} model 3 & 6\\
CeP1 & $\heartsuit$ & full and early & 0.7 & no \popthree{} stars, no adiabatic cooling period & 16\\
\hline
\end{tabular}
\end{table*}

\begin{table*}
\centering
\caption{IC naming convention.\label{table_ics}}
\begin{tabular}{|lccc|}
\hline
Mass and resolution code & DM particle mass ($10^3 \msol{}$) & gas particle mass ($10^3 \msol{}$) & softening length (pc) \\
\hline
M1L & 20.0 & 4.23 & 9.76\\
M1H & 5.0 & 1.06 & 6.15\\
M3L & 60.0 & 12.7 & 14.1\\
M3H & 15.0 & 3.17 & 8.87\\
M5L & 100.0 & 21.2 & 16.7\\
M5H & 25.0 & 5.29 & 10.5\\
M7L & 140.0 & 29.6 & 18.7\\
M7H & 35.0 & 7.40 & 11.8\\
M9L & 180.0 & 38.1 & 20.3\\
M9H & 45.0 & 9.52 & 12.8\\
\hline
\end{tabular}
\begin{tabular}{|lc|}
\hline
Rotation code & Physical velocity ($\kms{}$) \\
\hline
R00 & 0.0 \\
R05 & 5.0 \\
R10 & 10.0 \\
\hline
\end{tabular}
\end{table*}

\section{Analysis}\label{section_analysis}

\begin{figure*}
\centering{}
\includegraphics[width=\textwidth]{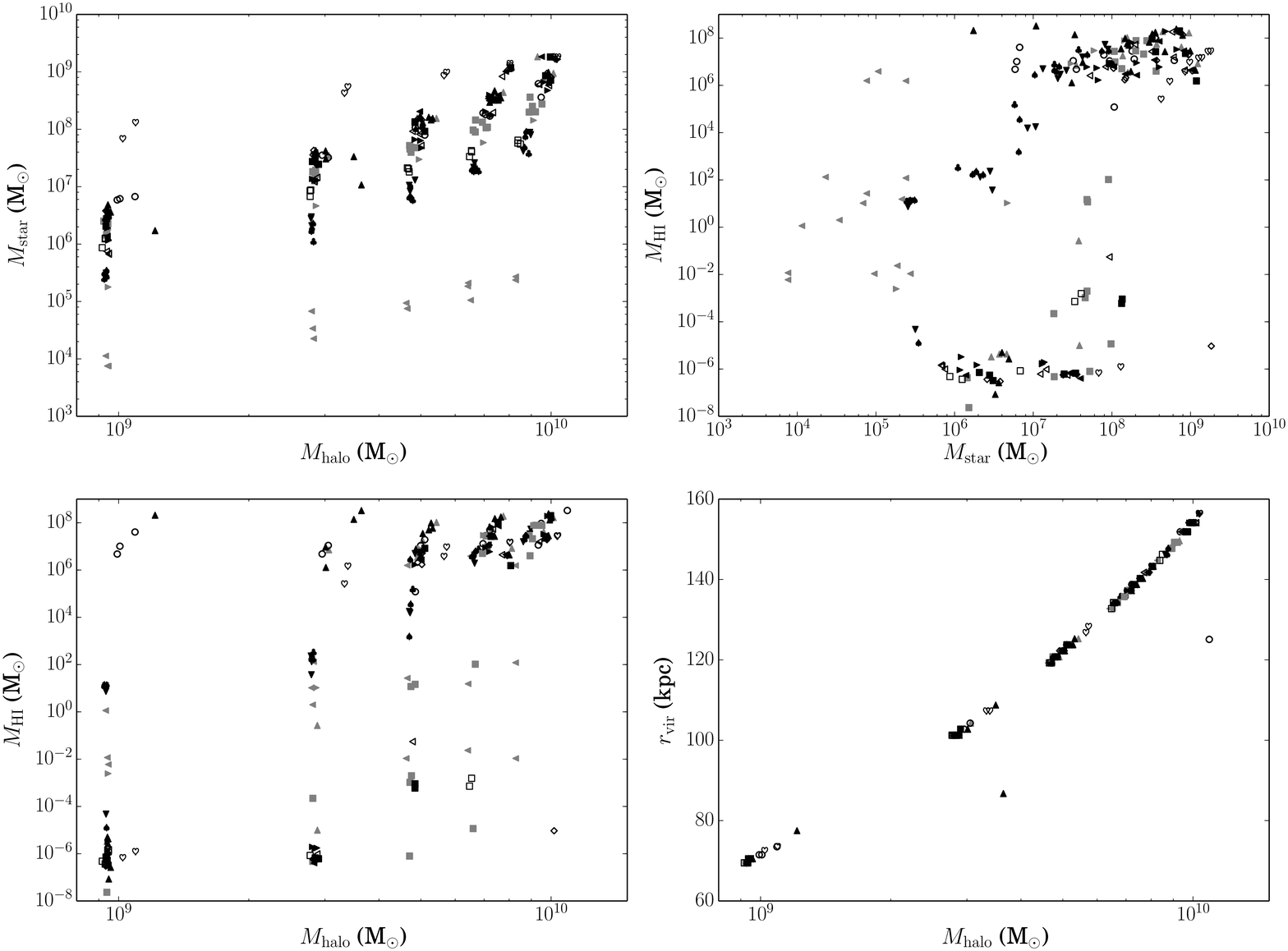}
\caption{Some general final properties of our simulations. \emph{Top left}~: stellar mass within the virial radius as a function of the total virial mass, \emph{top right}~: neutral gas mass within the virial radius as a function of the stellar mass within the virial radius, \emph{bottom left}~: neutral gas mass within the virial radius as a function of the total virial mass, \emph{bottom right}~: virial radius as a function of the total virial mass. The virial radius was calculated as the radius at which the mean density of the halo equals 200 times the mean density of the Universe. The symbols represent the different code and parameter values following \tableref{table_codes}. \label{fig_m_theory}}
\end{figure*}

Since we are interested in the location of our simulations on the BTFR, we need to determine (a) baryonic masses for all simulations, which are the sum of the stellar masses and neutral gas masses, and (b) circular velocities for all simulations. Since the BTFR is an observational relation, we will try to determine these quantities in the same way observers would do. We of course also have access to precise values for most of these quantities (some halo properties are shown in \figureref{fig_m_theory}), which will allow us to assess the validity of some observational proxies. Apart from the BTFR, we will also determine some other global galaxy properties and compare them to observed dwarf galaxies, as a further validation of our models. All analysis was performed using our open source analysis tool \hyplot{}\footnote{http://sourceforge.net/projects/hyplot/}. Since showing the results for all 250 simulations in the same plot makes the plots very hard to read, we also provide online interactive versions of some of our plots\footnote{http://www.dwarfs.ugent.be/btfr}.

\subsection{Stellar mass}\label{subsection_stellar_mass}
Numerically, we can determine the stellar mass by simply adding together the masses of the star particles in the simulation. However, the value obtained in this way will likely be larger than what will be observationally observed. Not all star particles reside within the central regions of the galaxy and will hence be observed. And even the star particles that are in the central regions might be too old and weak to be observationally detectable. To obtain mock observational values for our simulations, we fit a S\'{e}rsic profile to the surface brightness profile of our galaxy, cut off at a surface brightness of 30$\magarcsec{}$, and use this to estimate a half light radius ($R_e$) and an absolute magnitude in the V- and I-band. Surface brightnesses are estimated from the age and metallicity of the star particles using the tables of \cite{vazdekis}. We then use the I-band luminosity and V-I colour to estimate the total stellar mass \citep{bell}.

Not all our simulations contain enough stars to fit a general S\'{e}rsic profile. We therefore use a simple exponential curve if the S\'{e}rsic profile is visibly a bad fit. All simulations of the C7P6 model have very little stars, which are spread out over huge volumes, so that even an exponential fit is impossible. For these models, we will use the sum of the star particle masses as stellar mass, keeping in mind that these systems in no way resemble real dwarf galaxies. Some of the more massive models (mainly with the M9R00L and M9R00H initial conditions) have very steep S\'{e}rsic profiles with indices larger than 1.5. The total magnitudes obtained from these fits are generally much smaller than the total magnitudes obtained by summing the luminosities of the star particles, so that the estimated mass for these systems tends to be wrong by a large factor. We will therefore also fit an exponential curve to the central brightness profile of these models, so that the estimated total magnitude better resembles the summed luminosities. In total, 216 of the 250 simulations were fitted with a general S\'{e}rsic profile and 18 were fitted using an exponential profile. All 16 simulations of the C7P6 models were discarded.

\begin{figure}
\centering
\includegraphics[width=0.5\textwidth]{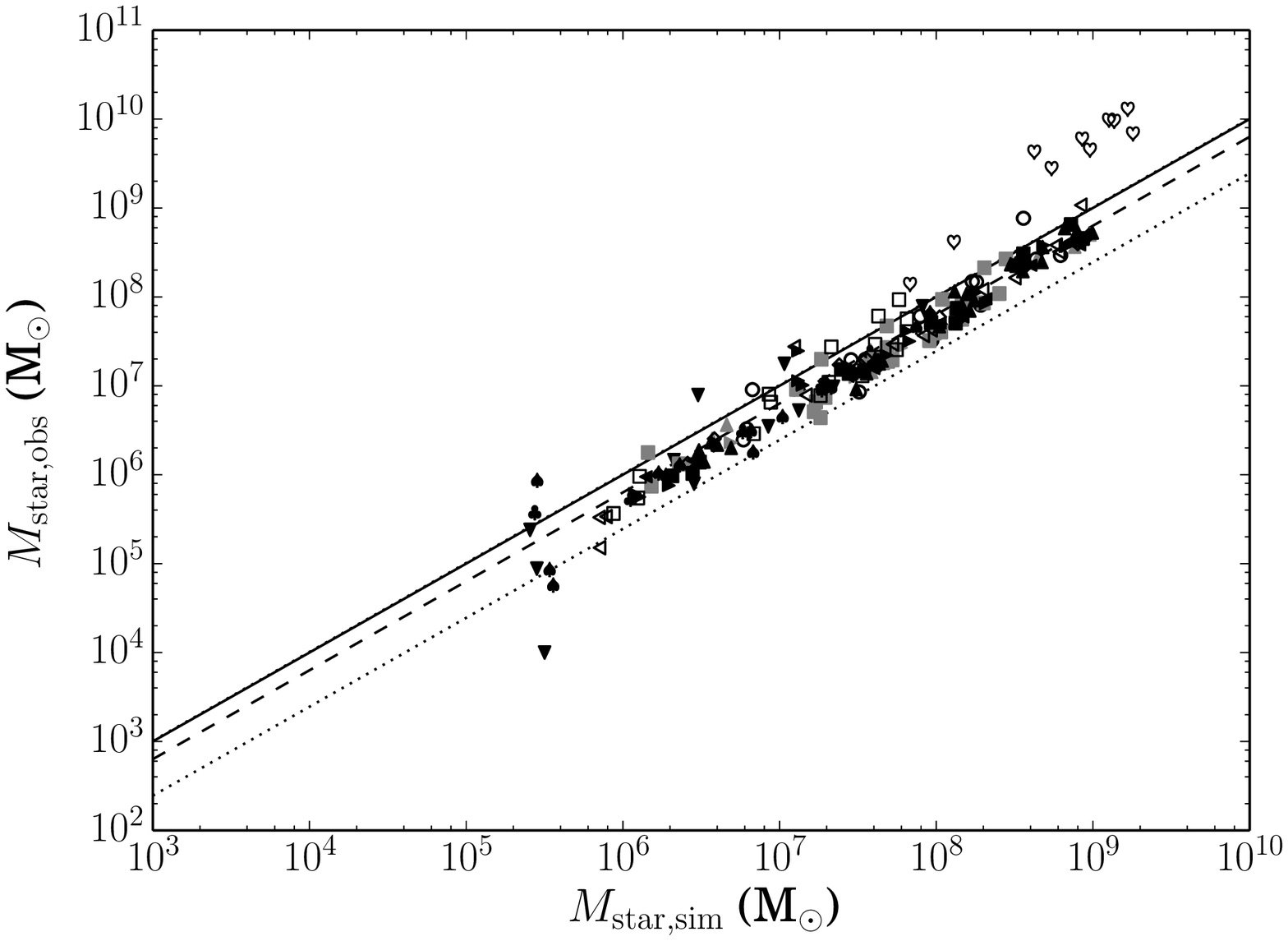}
\caption{The mock observational stellar mass as a function of the actual value. The different symbols represent the different code and parameter values, following \tableref{table_codes}. The dashed line represents the mean ratio of observed and simulated masses and the dotted lines are a 1$\sigma{}$ interval around these values. The full line is a 1:1 relation. Model C7P6 (gray left pointing triangles) lies on the full line since we set its mock observational mass to be equal to the actual value.\label{fig_Mstar}}
\end{figure}

In \figureref{fig_Mstar}, we show the actual stellar mass as a function of the mock observational value. We find quite a good correlation between both values, with a mean observed to simulated stellar mass ratio of $0.63\pm{}0.38$ (discarding the C7P6 and CeP1 models and models with a S\'{e}rsic index larger than 1.5), which means this simple technique indeed works. The fact that we systematically underestimate the stellar mass is due to the differences in assumed IMF between the models used to derive luminosities for our star particles \citep{vazdekis} and the mass to light ratio models of \cite{bell}. 

\subsection{Neutral gas mass}
As described above, we need a detailed model of the ionization state of the gas to correctly describe the hydrodynamics of the gas in our simulation. To this end, we keep track of two metallicity tracers and the temperature, density and redshift of the SPH particles and determine the ionization equilibrium by a 5D interpolation on pre-calculated tables \citep{sven_cooling_curves}. We can use the same tables used in the simulation to calculate the neutral fraction of the gas for the snapshots of the simulation in post-processing. By multiplying these neutral fractions with the masses of the gas particles, we can hence very easily determine neutral gas masses.

As for the stellar mass, these numerically determined masses might overestimate the observational values, since not all gas will be confined to the central regions of the galaxy. However, observational H\,\textsc{i} clouds can be significantly larger than the physical size of the galaxy as estimated from $R_e$.

Since we consider galaxies in isolation and in the presence of an ionizing UVB, we do not expect neutral gas at large distances from the galactic halo. So rather than introducing an arbitrary cutoff on the gas taken into account for the neutral gas mass, we will always consider all gas.

\subsection{Circular velocity}

\begin{figure*}
\centering{}
\includegraphics[width=\textwidth]{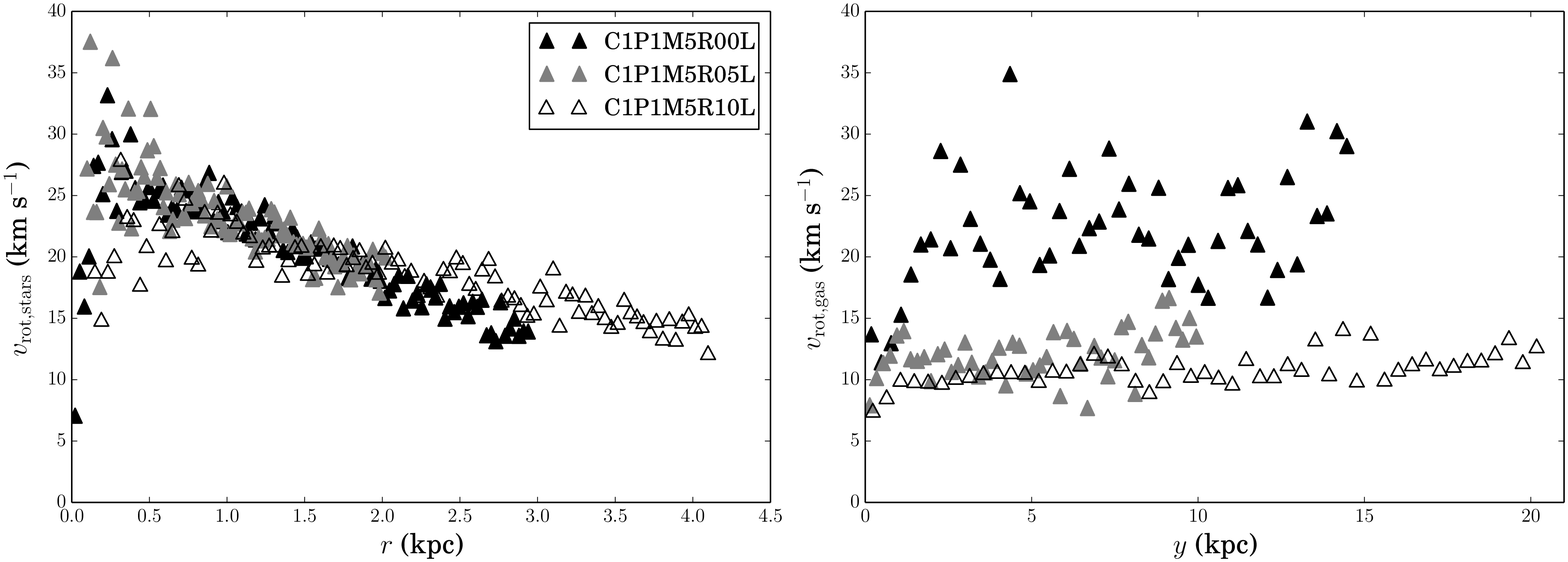}
\caption{The rotation velocity of the stars and gas in three simulations using the same model, but with initial conditions with different rotation parameters. \emph{Left}~: rotation velocity of the stars, computed as their tangential velocity and weighted with the number of RGB stars per particle, cut off at twice the estimated half light radius. \emph{Right}~: rotation velocity of the gas, computed as the velocity in the $x$-direction and measured as a function of the distance to the $y$ axis (the $z$ axis corresponds to the rotation axis for models receiving an initial rotation), cut off at ten times the estimated half light radius. The symbols represent the different rotation parameters, as indicated in the legend.\label{fig_vprof}}
\end{figure*}

The circular velocity of a halo is a measure for the strength of its gravitational potential. Theoretically, one can construct a circular velocity profile of a galaxy based on the gravitational potential inferred from its total mass profile. To characterize this profile, one then has to either choose a characteristic radius at which to evaluate this profile or take e.g. the maximal value as ``the'' circular velocity. The latter makes sense for halos with a strictly rising circular velocity profile that tends to become flat at larger radii, and for which both methods should obtain the same result if the chosen radius is large enough.

Observationally, the total mass profile is not accessible and one has to derive the circular velocity using a tracer, e.g. the neutral gas halo or the stellar body, see \figureref{fig_vprof}. The BTFR of \citet{mcgaugh} uses the maximal circular velocity as obtained from resolved H\,\textsc{i} rotation profiles and we will try to use the same definition for the circular velocity of our galaxies whenever possible. Alternatively, \citet{mcgaugh} also uses circular velocities derived from line widths if a full rotation curve is not available. We will use the same technique and compare it with the circular velocities obtained from rotation curves.

\begin{figure}
\centering
\includegraphics[width=0.5\textwidth]{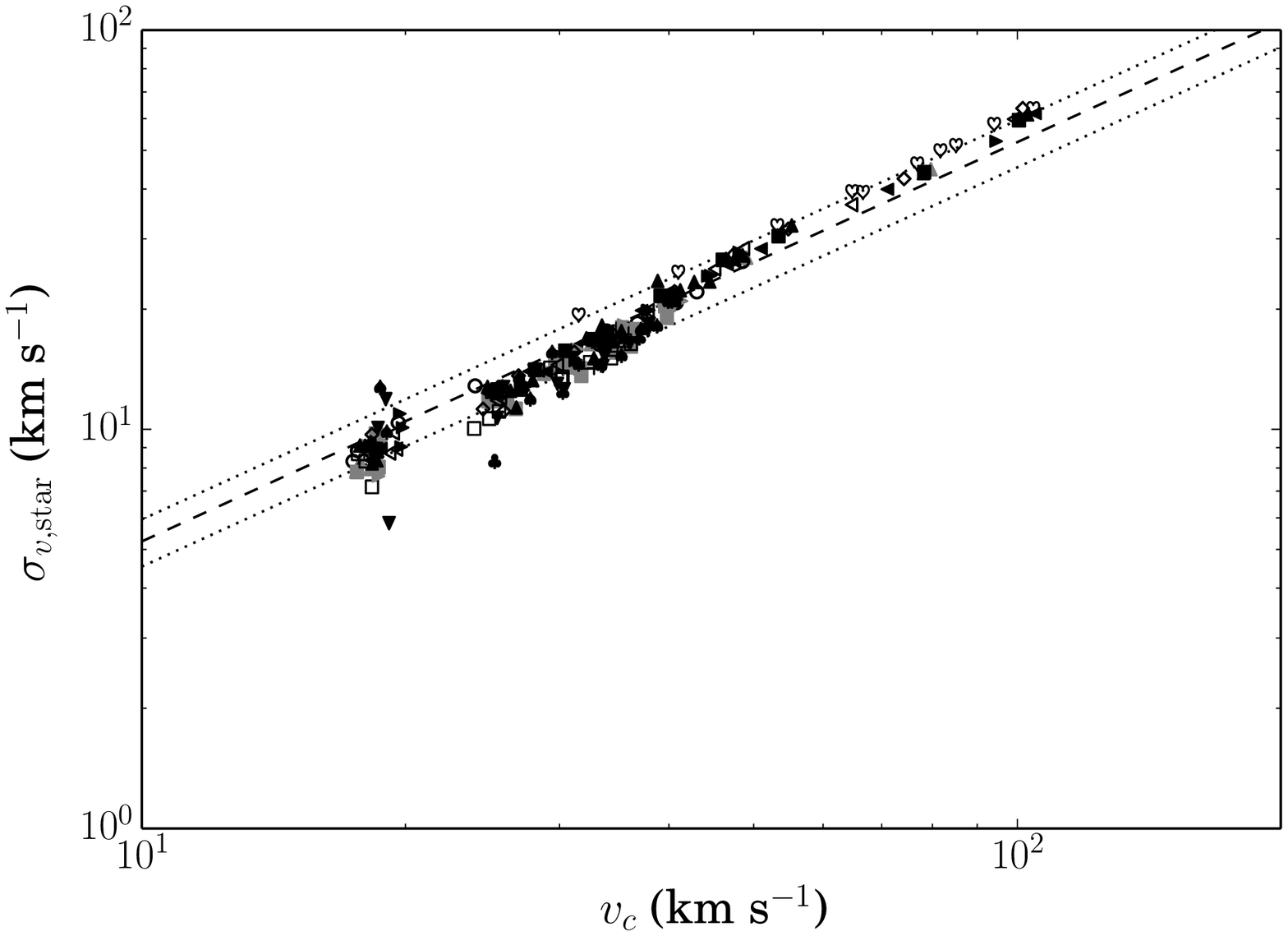}
\caption{The velocity dispersions of the stars within a sphere with radius two times the half light radius, weighted with the number of RGB stars per star particle, as a function of the theoretical circular velocity. The symbols represent the different code and parameter values following \tableref{table_codes}, the dashed line represents a linear least-squares fit to the points, and the dotted lines represent a 1$\sigma{}$ interval around this fit.\label{fig_vstar}}
\end{figure}

Unlike \citet{mcgaugh} however, we cannot use H\,\textsc{i} for all our simulations, since not all our simulations contain neutral gas at the end. We therefore will introduce an alternative method to estimate the circular velocity of our halos based on the velocity profile of the stars. To this end, we consider an observationally accessible property of the stellar body, namely the velocity dispersion along a line of sight. For an isothermal sphere, the stellar velocity dispersion can be shown to correlate linearly with the circular velocity \citep{binney_tremaine}. In \figureref{fig_vstar}, we show the velocity dispersions in the $x$, $y$ and $z$ direction of the stars within a sphere with radius $2R_e$ for all our simulations as a function of the actual theoretically determined maximal circular velocity. To correct for the fact that not all stars are equally bright and contribute equally to the spectra that can be used to observationally measure velocity dispersions, we have weighted the contributions of the different star particles with the number of Red Giant Branch (RGB) stars per particle. Apart from a few outliers which do not contain enough stars to fit a S\'{e}rsic profile and for which the whole procedure is in fact meaningless, the values clearly trace out a linear relation.

A least-squares fit to the simulation data yields the following relation between the ``observed'' stellar velocity dispersion and the circular velocity of the halo~:
\begin{equation}
\sigma_v = \left(0.52 \pm 0.07 \right) v_\smalltext{c}.
\end{equation}
If no neutral gas is available, we will derive the circular velocity using this relation. We could of course also use the theoretical value, but this is in no way observationally accessible, while the proposed relation in principle is.

\begin{figure}
\centering
\includegraphics[width=0.5\textwidth]{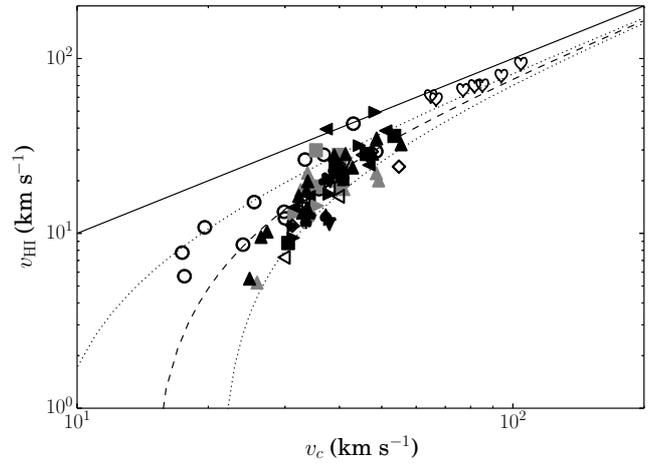}
\caption{The circular velocity derived from the rotation profile of the neutral gas as a function of the theoretical circular velocity. The symbols represent the different code and parameter values following \tableref{table_codes}, the full line represents a 1:1 relation, the dashed line is a linear least-squares fit, while the dotted lines correspond to a 1$\sigma$ interval around this fit. Only the simulations for which a workable rotation profile was found are shown.\label{fig_vHI}}
\end{figure}

If neutral gas is available, we can of course use the rotation of the neutral gas as a tracer for the circular velocity. To this end, we produce a mock rotation curve of the neutral gas and determine its maximum value. We visually checked all rotation curves and only kept the values that were obtained from curves with a clear rotation and enough neutral gas. In \figureref{fig_vHI} we show the circular velocity derived from the neutral gas as a function of the theoretical value. The agreement is reasonable on the massive end, but gets worse for halos with a small theoretical circular velocity. Overall, we find that H\,\textsc{i} rotation curves underestimate the true circular velocity.

\begin{figure}
\centering
\includegraphics[width=0.5\textwidth]{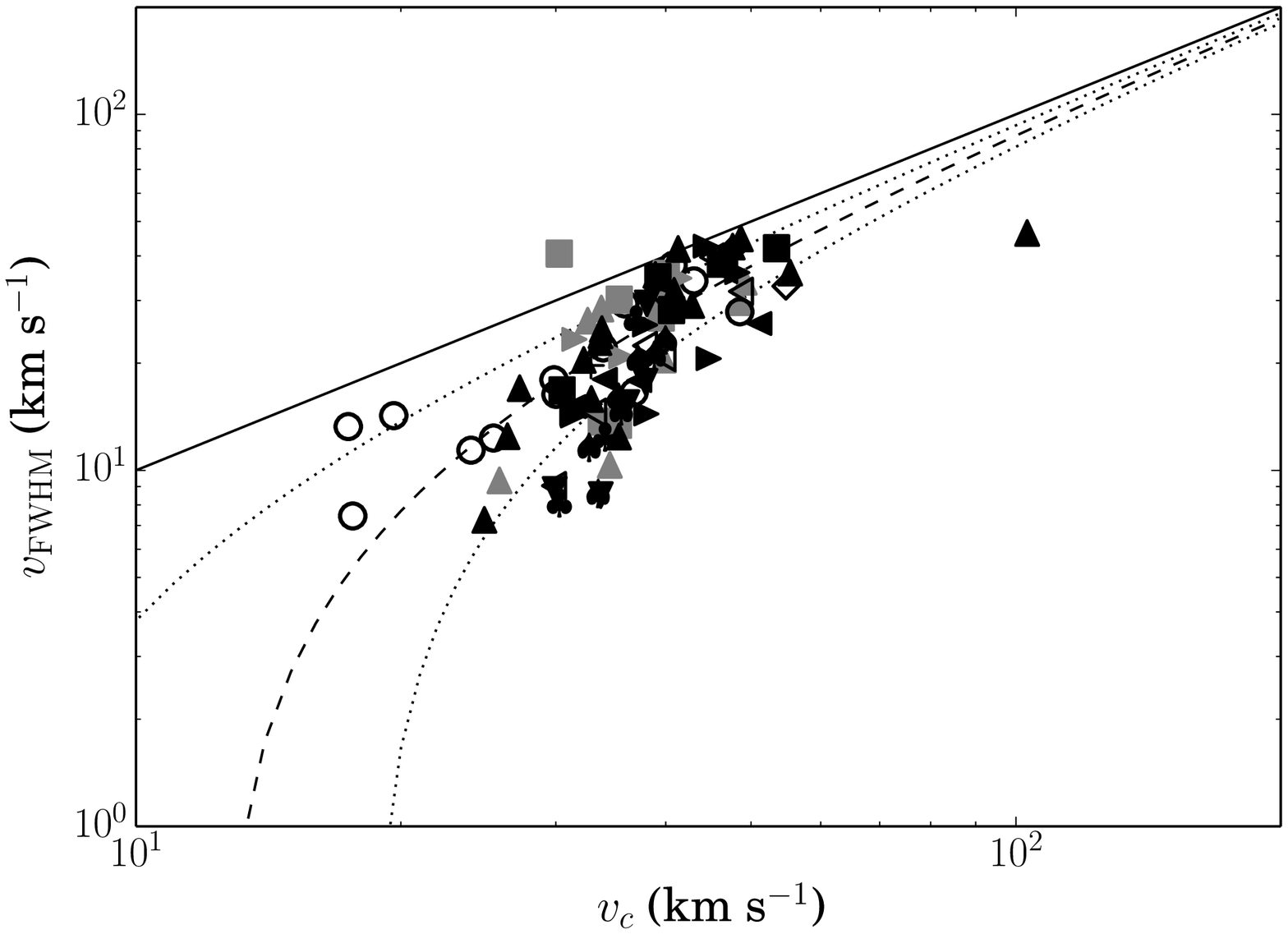}
\caption{The circular velocities obtained by using line widths as a function of the theoretical value. The symbols represent the different code and parameter values following \tableref{table_codes}, the full line represents a 1:1 relation, the dashed line is a linear least-squares fit and the dotted lines a 1$\sigma$ interval around this fit. Only the simulations for which a workable mock spectrum could be constructed are shown.\label{fig_vFWHM}}
\end{figure}

\begin{figure}
\centering
\includegraphics[width=0.5\textwidth]{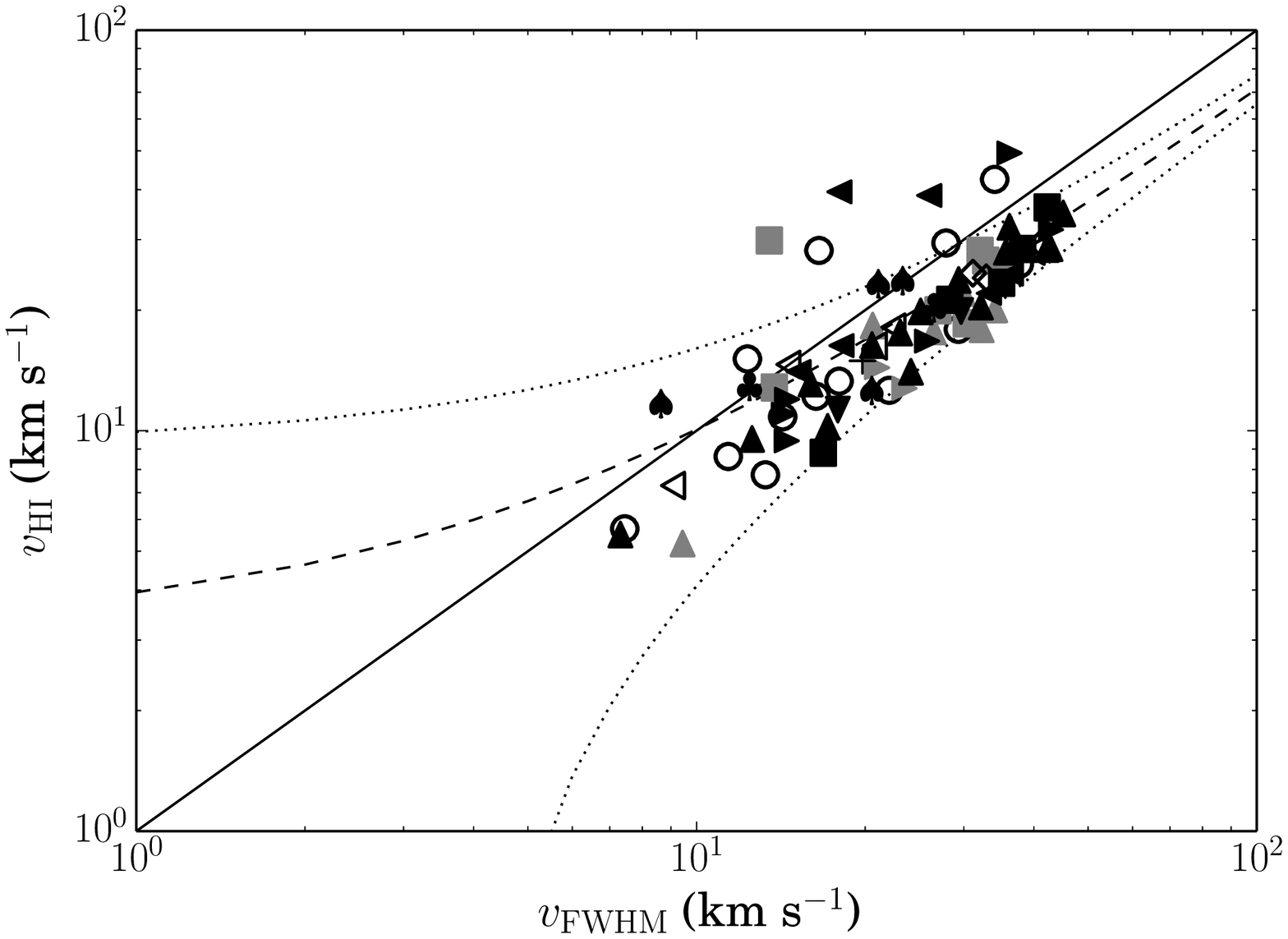}
\caption{The circular velocities obtained by using full rotation curves as a function of the ones obtained by using line widths of mock spectra. The symbols represent the different code and parameter values following \tableref{table_codes}, the full line represents a 1:1 relation, the dashed line is a linear least-squares fit and the dotted lines a 1$\sigma$ interval around this fit. Only simulations for which both a workable rotation curve and a workable mock spectrum were constructed are shown.\label{fig_vFWHM_vs_vHI}}
\end{figure}

\figureref{fig_vFWHM} shows the circular velocities obtained by using line widths for the neutral gas, following \citet{mcgaugh}, as a function of the theoretical value. To produce these values, we fitted normal distributions to mock spectral lines. The circular velocity is then half $W_{20}$, the width of the Gaussian bell curve at 20 per cent of its maximum value. We average this value over 8 distinct edge-on lines of sight and again only kept the values for which we could visually confirm a decent fit. We find a slightly better agreement than for the velocities derived from rotation curves, but the theoretical value is still underestimated for most of the simulations.

\figureref{fig_vFWHM_vs_vHI} shows the same velocities, but as a function of the values obtained using the rotation curves. We notice that the values derived from the line widths are systematically higher than the values obtained from the rotation curves. Not all simulations have circular velocities determined using both methods (because of a bad quality rotation curve or a bad Gaussian fit to the line widths), in which case we set the corresponding velocity to zero. In total, 113 simulations have enough neutral gas to estimate a circular velocity from the gas. For 90, we were able to determine the circular velocity using both methods. 11 more have circular velocities derived from the rotation curve of the gas, so that in total we will use rotation curve based circular velocities for 101 simulations. For the remaining 12 simulations with enough neutral gas, but for which the rotation curves could not be used, we will use the value derived from line widths instead.

It should be noted that observed circular velocities sometimes correct for turbulence by adding correction factors to the quadratic circular velocities, both for those estimated from rotation profiles as from line widths. For an exponential disk, it is straightforward to determine an asymmetric drift correction. However, our galaxies do not show exponential disks, which makes it much harder to estimate correction factors. Moreover, as noted by \cite{mcgaugh}, these corrections are typically small for low mass halos. We will therefore quote the circular velocity derived from the neutral gas as a directly observable quantity, without applying any corrections to it.

\subsection{BTFR}

\begin{figure}
\centering
\includegraphics[width=0.5\textwidth]{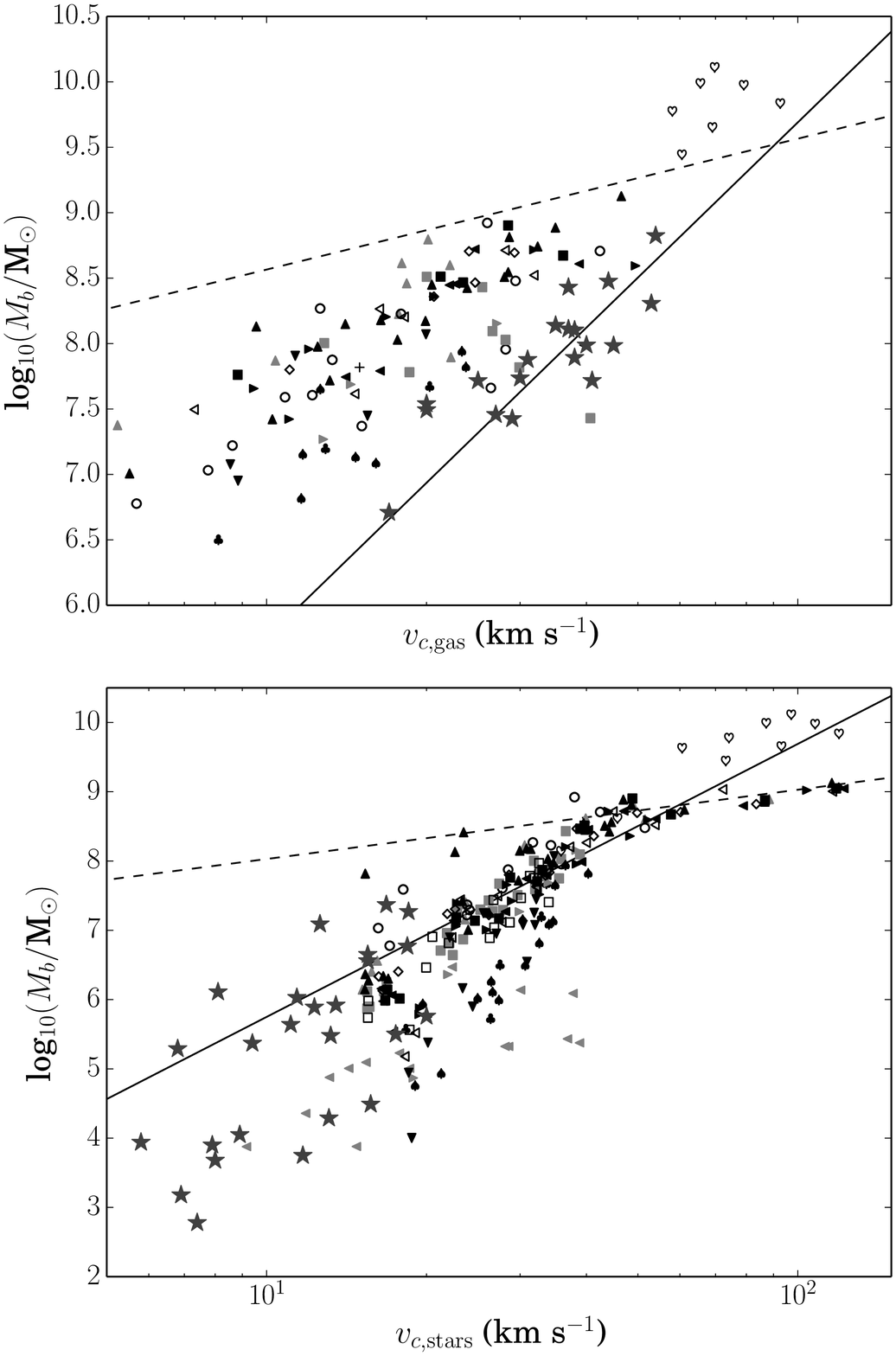}
\caption{The BTFR for all 250 simulations in the set. The top panel shows the BTFR with circular velocities derived from the neutral gas, the bottom panel those derived from the stellar velocity dispersion. The symbols represent the different code and parameter values following \tableref{table_codes}, the stars are the observational values from \citet{mcgaugh} and \citet{mcgaugh2010}. The full line is the fit of \citet{mcgaugh}, while the dashed lines are least-squares fits to the simulation data.\label{fig_btfr_all}}
\end{figure}

\figureref{fig_btfr_all} shows the BTFR for all our simulations. The top panel shows the BTFR obtained using circular velocities derived from the gas, while the bottom panel shows those derived from the stellar velocity dispersion. We have also indicated the power law fit of \citet{mcgaugh} in both panels. It is clear that (a) the BTFR with velocities from the gas lies above the observational relation for the bulk of our simulations, and (b) there is a clear difference between the relation obtained using gas circular velocities and stellar circular velocities. Since we showed above that stellar circular velocities are more closely related to the real theoretical circular velocities, this means that circular velocities derived from neutral gas observations systematically underestimate the circular velocity. Notice also that there is a bend in the bottom BTFR at stellar circular velocities of $\sim{}30\kms{}$, which indicates the transition between a mass regime where all halos keep gas and form stars for the entire lifetime of the simulation, and a mass regime where the halos loose their gas and contain only old stars. This bend is absent in the top panel, since we cannot derive circular velocities from the neutral gas for the halos which loose their gas. The observational BTFR of \citet{mcgaugh2010} does not show this bend, but has considerably more scatter in this regime than the observed BTFR of \citet{mcgaugh}.

Keeping this in mind, we will always show two BTFRs when discussing our simulations~: the one derived from the gas, which we can directly compare to the observations of \citet{mcgaugh}, but which can only be calculated if there is neutral gas, and the one derived from the stellar velocity dispersions, which we can calculate for all our simulations, and which we will compare to the observed BTFR of \citet{mcgaugh2010}. To guide the eye, we will compare the latter to a least-squares fit to the BTFR in the bottom panel of \figureref{fig_btfr_all}, which will tell us how the particular model compares to the other models in the set.

\subsection{Star formation histories}
The final stellar mass and neutral gas mass of a simulated galaxy are completely determined by its star formation history (SFH). Differences in these masses between different models will therefore always trace back to differences in the SFH for these models and it is important to have a good way of investigating the SFH. Observationally, the SFH is inferred from the colour magnitude diagram (CMD), using a complex fit of stellar population models with different ages and metallicities. Since there is a degeneracy in colour between old metal rich and young metal poor stars, there is a large uncertainty on these observed SFHs, which makes it difficult to compare them with our simulations. Moreover, these observations suffer from the same luminosity constraints discussed above and only trace the central galaxy.

To explain the differences between our models, we need a more detailed SFH than what is observationally accessible and we need to take into account all stars, not only those that happen to be in the central galaxy at the end of the simulation. We will therefore use a theoretical SFH in our analysis, shown as the effective star formation rate (SFR) as a function of time. It is determined by counting the stellar particles that are born in a specific time interval during the simulation, irrespective of their position or intrinsic properties. We note that these SFHs cannot be compared to the observed SFHs in e.g. \citet{weiszA}, which are based on a fit to the observed CMD. Determining mock observational SFHs based on a fit to mock CMDs falls outside the scope of this paper and will be subject of future work.

\subsection{Scaling relations}
To check whether our simulated galaxies look like observed galaxies, we look at the so-called \emph{scaling relations}. These relations describe the correlations of various global observational quantities that were found for observed dwarf galaxies and they loosely define the typical size and luminosity for a dwarf galaxy. \citet{annelies_sfp} showed that there is a large range of stellar feedback parameters which gives rise to dwarf galaxies well within the range of these scaling relations, so we cannot in general use them to constrain our models. However, they do provide a first check on the result, as models that do not lie on the scaling relations cannot be considered to represent real dwarf galaxies.

\begin{figure*}
\centering
\includegraphics[width=\textwidth]{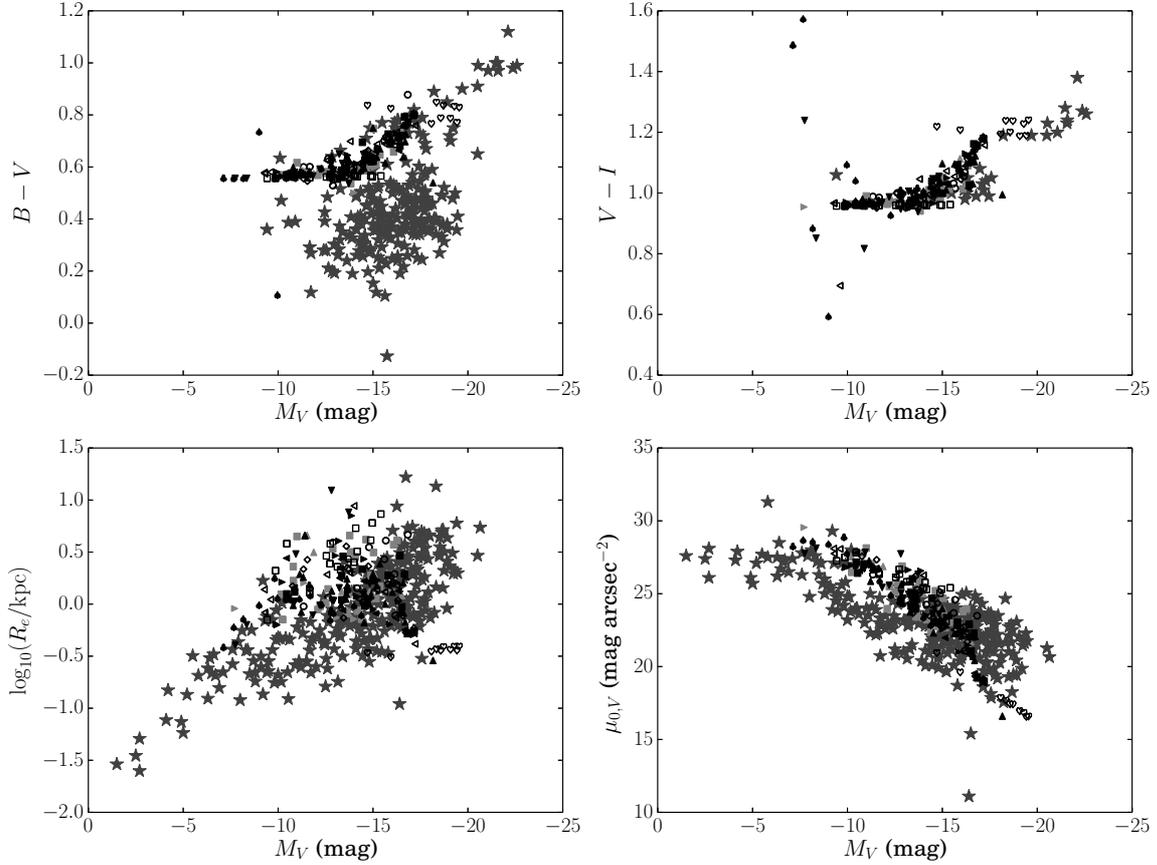}
\caption{The position of our models on four observational scaling relations. \emph{Top left}~: $B-V$ colour, \emph{top right}~: $V-I$ colour, \emph{bottom left}~: half-light radius, \emph{bottom right}~: central surface brightness. The symbols represent the different code and parameter values following \tableref{table_codes}, the stars are the observational values from \citet{vanzee2000}, \citet{magorrian}, \citet{geha}, \citet{grebel}, \citet{vanzee2004}, \citet{hunter}, \citet{dunn}, \citet{mcconnachie}, \citet{mcquinn}, \citet{rhode}, and \citet{tollerud}.\label{scaling_relations}}
\end{figure*}

The scaling relations are shown in \figureref{scaling_relations}. The observational data with which we compare consists of early and late type galaxies within the Local Volume \citep*{vanzee2000,grebel,hunter,mcconnachie}, including recent additions like Leo P \citep*{mcquinn, rhode}, and Pisces A and B \citep*{tollerud}, galaxies within the Coma \citep*{graham}, Virgo \citep*{vanzee2004} and M81 cluster \citep*{lianou}, and isolated dwarf galaxies \citep*{vanzee2000, magorrian, geha, grebel, hunter, dunn}.

Model C7P6 is not shown, since we were unable to fit a S\'{e}rsic profile to these simulations (see \sectionref{subsection_stellar_mass}). Simulation CaPaM1R00L is also not shown, since for that simulation the B-band S\'{e}rsic fit was very bad (leading to an unrealistic $B-V$ colour of -4.96). The large scatter at the high surface brightness end of the $V-I$ colour is similarly caused by bad quality S\'{e}rsic fits to the low stellar mass simulations at this end, but these simulations are shown. All other simulations fall well within the observational scatter.

We do notice two important effects~: all simulations have $B-V$ colours at the high end of the observational distribution, and all simulations have large half light radii. Both effects can be linked to a large initial peak in the galactic star formation, as this will (a) produce a predominantly old stellar population, and (b) lead to a high stellar feedback at the start of the simulation that will disperse the interstellar gas and lead to an overall larger and shallower halo.

\subsection{Metallicities}
To determine stellar metallicities, we follow \citet{kirby}, who observed the stellar metallicities for 7 local group dIrrs from the optical colours of its red giant branch (RGB). To this end, we use the stellar evolution tracks of \citet{bertelli_2008, bertelli_2009} for \popone{} and \poptwo{} stars and those of \citet{marigo} for \popthree{} stars to estimate the fraction of the SSP represented by a star particle that will reside on the RGB at any given time. We then weigh the metallicities of the star particles within the galaxy with this fraction to calculate the average [Fe/H]-metallicity.

\begin{figure}
\centering
\includegraphics[width=0.5\textwidth]{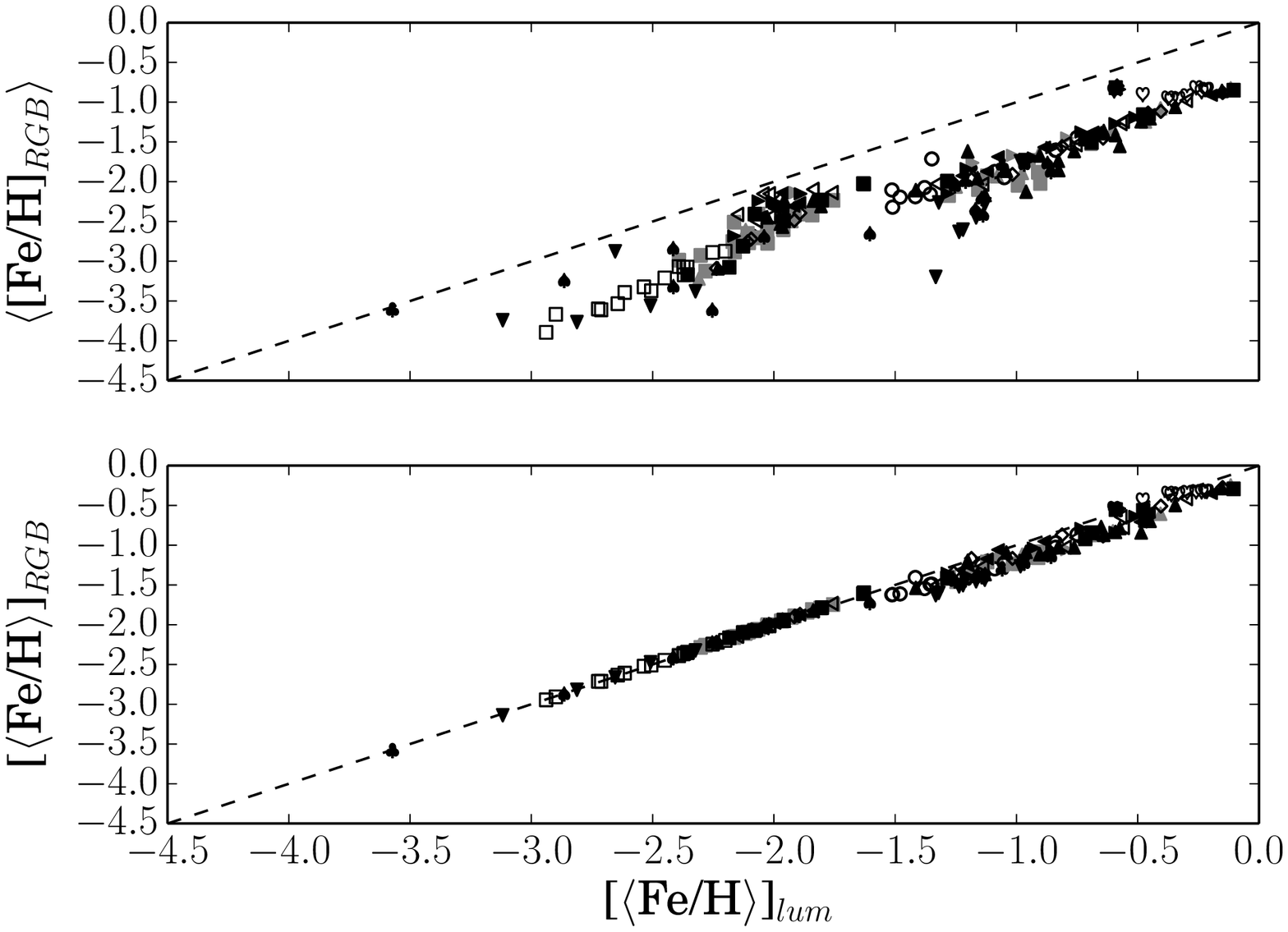}
\caption{The metallicities as determined from the RGB colours as a function of the spectroscopic luminosity weighted metallicities for all simulations. The upper panel shows the mean RGB metallicity as used in \citet{kirby}, the bottom panel shows the metallicity calculated from the mean metal content of the RGB stars. The symbols represent the different code and parameter values following \tableref{table_codes}, the dashed lines represent a 1:1 relation. Simulations which only formed a single generation of \popthree{} stars (model C7P6) and hence have no metallicity ([Fe/H] = -99.0) are not shown.\label{fig_zlum_vs_zRGB}}
\end{figure}

We note that this method will yield lower metallicities than for example a luminosity weighted average metallicity as can be obtained from stellar spectroscopy, since it is biased towards the older star particles, that will have significantly higher RGB fractions. This effect is visible for the high metallicity end of \figureref{fig_zlum_vs_zRGB}. The low metallicity end corresponds to galaxies with a predominantly old stellar population, where both methods yield the same metallicities.

We also compare the average $\langle$[Fe/H]$_\smalltext{RGB}\rangle$ metallicities with the metallicities of the averaged metal content of the galaxies, [$\langle$Fe/H$_\smalltext{RGB}\rangle$]. The former is the quantity discussed in \citet{kirby}, while the latter is a more meaningful measure of the total metallicity of the galaxy. It is clear from \figureref{fig_zlum_vs_zRGB} that $\langle$[Fe/H]$_\smalltext{RGB}\rangle$ underestimates the metallicity of the galaxy by giving too much weight to low metallicity RGB stars.

\begin{figure}
\centering
\includegraphics[width=0.5\textwidth]{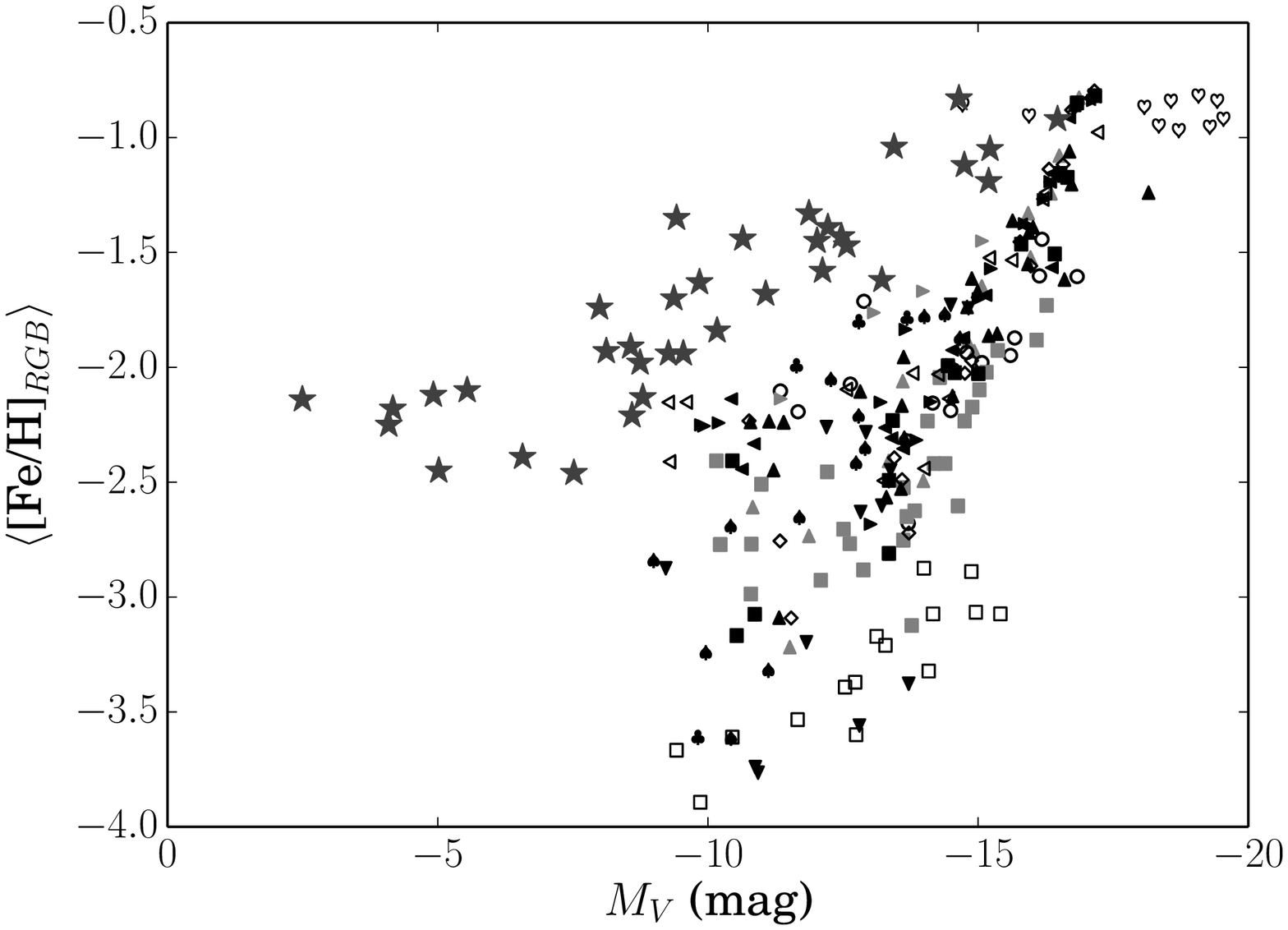}
\caption{The metallicities of all simulations as a function of their $V$-band magnitude. The symbols represent the different code and parameter values following \tableref{table_codes}, the stars are the observational values from \citet{kirby}.\label{fig_metallicities}}
\end{figure}

\figureref{fig_metallicities} shows the average $\langle$[Fe/H]$_\smalltext{RGB}\rangle$ metallicities as a function of the $V$-band magnitudes for all simulations, compared to the observational data of \citet{kirby}. It is clear that all our models lie below the observations and our too metal poor. The reason for this is that most of our simulations form most of their stars early on and then stop forming stars or form very little stars for the rest of the simulation. Most stars are hence born from gas that is metal poor. In \sectionref{section_results} we will discuss ways to reduce the relative strength of this initial star formation peak. It is already worth noting that the models that somewhat succeed in this are closer to the observed metallicities.

\section{Results and discussion}\label{section_results}

\subsection{Stochasticity}\label{subsection_stochasticity}
Before we can start comparing different models, we first have to assess the effect of small random changes in the system on the outcome of the simulations. This will help us distinguish between differences caused by stochastic effects and fundamental differences caused by physical ingredients. When discussing differences, we will focus on the BTFR and related quantities~: stellar and neutral gas masses, circular velocities and star formation histories.

\begin{figure}
\centering
\includegraphics[width=0.5\textwidth]{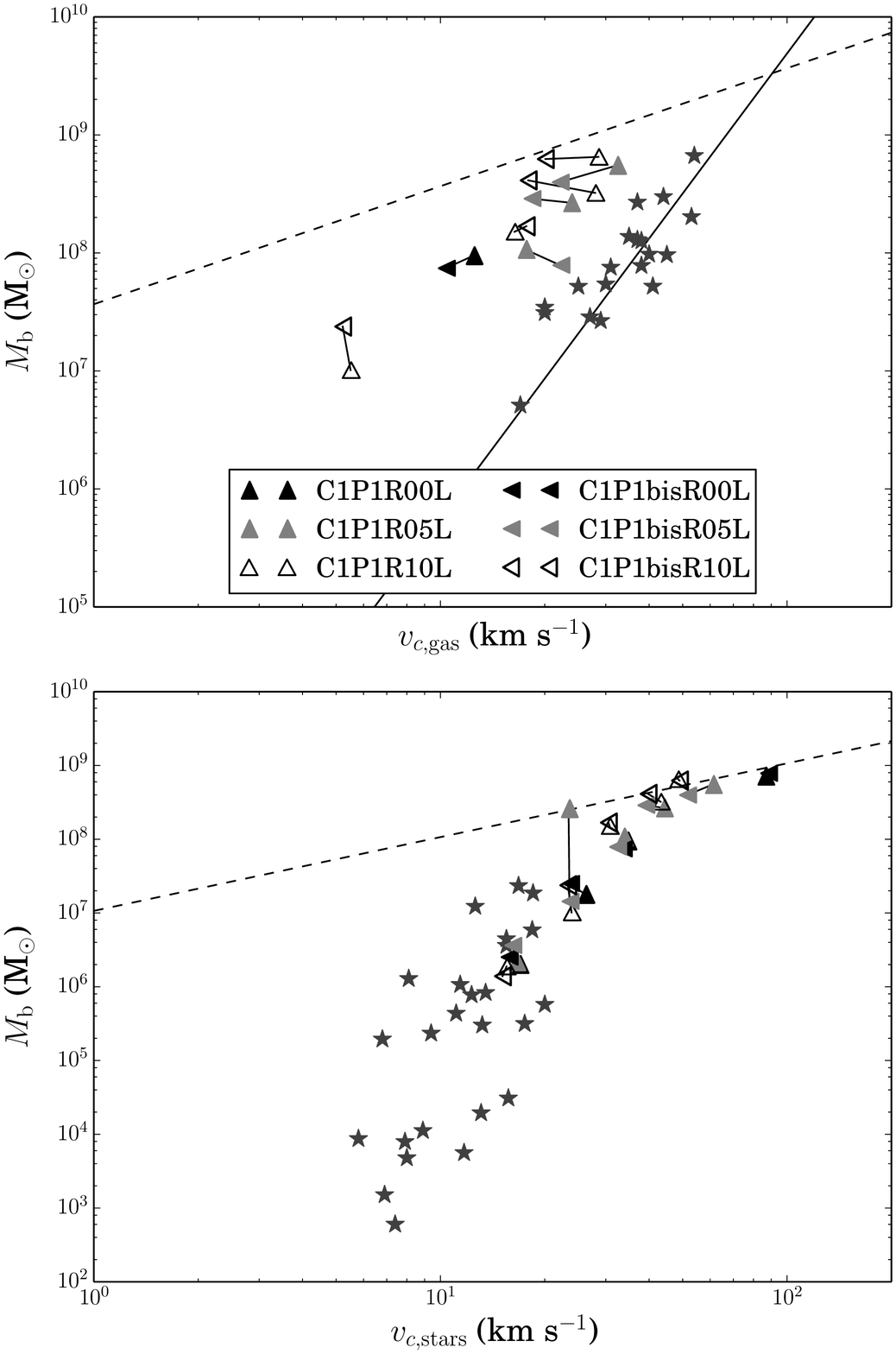}
\caption{The BTFR for the simulations using the same model, but with different initial conditions. The stars and the full line represent the observational data and a fit to it from \citet{mcgaugh}, the other symbols are our models, as indicated in the legend. The dashed lines correspond to the least-squares fit to all simulations in \figureref{fig_btfr_all}. For clarity, simulations that represent the same initial condition have been joined by a full line.\label{fig_stochasticity_btfr}}
\end{figure}

\figureref{fig_stochasticity_btfr} shows the BTFR for the same models, but with ICs generated with different random seeds (the C1P1bis model). Both BTFRs are clearly very similar, although there are small differences in final masses between the models with the same initial mass and rotation parameters. However, most models have almost no neutral gas at the end of the simulation, so the agreement stems mostly from very similar final stellar masses.

\begin{figure}
\centering
\includegraphics[width=0.5\textwidth]{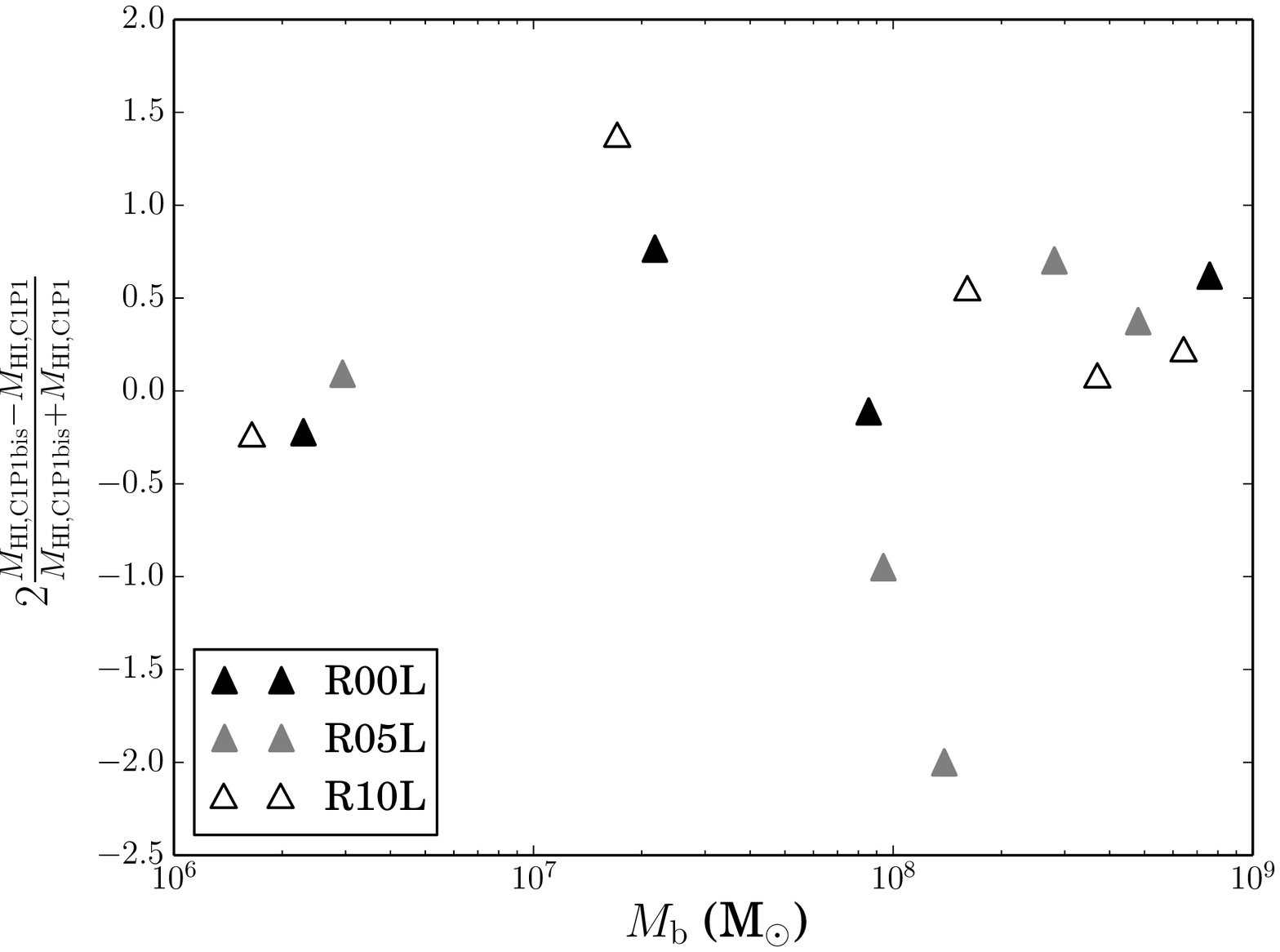}
\caption{The relative difference of the final neutral gas mass as a function of the average final baryonic mass for the simulations using the same model, but with differently sampled initial conditions. The symbols correspond to the different initial condition parameters as indicated in the legend.\label{fig_stochasticity_gas}}
\end{figure}

When we look at the final neutral gas masses of the different models \figurerefp{fig_stochasticity_gas}, we see the agreement between the models with different ICs is a lot worse, especially in the cases with intermediate neutral gas masses, where the relative difference can be up to a factor 2. This makes sense, since in these cases the galaxy has lost almost all of its gas by the end of the simulation, so the fraction that is left will be largely determined by small stochastic differences between the simulations.

\begin{figure}
\centering
\includegraphics[width=0.5\textwidth]{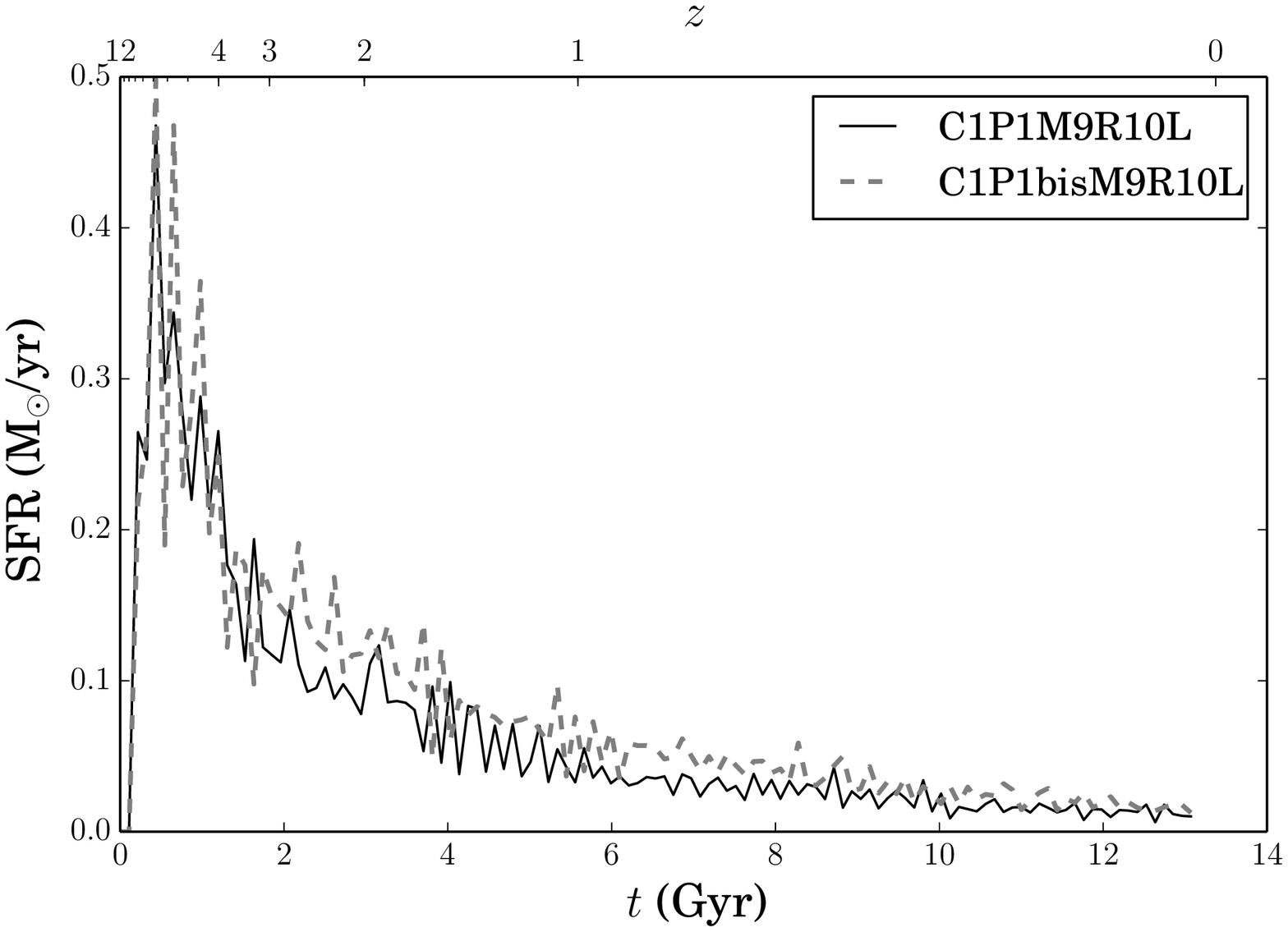}
\caption{The star formation history for two simulations using the same model, but with different initial conditions.\label{fig_stochasticity_sfr}}
\end{figure}

\figureref{fig_stochasticity_sfr} shows the SFH for the same model, with two different ICs. We see a good qualitative agreement over the entire run time of the simulation.

We conclude that stochastic effects do not significantly change the position of our models on the BTFR, nor their final stellar masses and SFHs. When discussing the neutral gas masses of different models however, we have to keep in mind the large sensitivity of these masses to stochastic effects, especially when there is neutral gas left but its final mass is low. This poses no real threat to our work, since the ultimate goal is to produce dwarf galaxies with final neutral gas masses that are of the same order of magnitude as the final stellar masses, in which case the outcome is less determined by stochastic effects. And for other models, our simulations are capable of determining whether or not a galaxy will be able to keep a significant fraction of its neutral gas, which is all that matters for our purposes.

\subsection{Convergence}\label{subsection_convergence}
Since we want to explore a large parameter range, we want to keep the computational cost of the simulations to a minimum. It is therefore important to carefully select the numerical resolution of the simulations. A high resolution will improve the accuracy of the results, but will at the same time increase the run time of the simulation. Meanwhile, our sub-grid model is based on the assumption that we can treat individual star particles in our simulations as stellar populations with statistically averaged properties, which limits our highest resolution to a star particle mass of $\sim 1000\msol{}$.

We therefore opt to use lower resolution simulations with shorter run time for the bulk of our simulations, as long as this resolution is sufficient to resolve the properties of interest. The simulations in e.g. \citet{joeri}, \citet{annelies_mergers}, and \citet{robbert} use 200,000 SPH particles, which for the lowest mass simulations in our mass range of interest roughly corresponds to five times the minimal star particle mass. The mean run time of these simulations is however too long for a large parameter study. Simulations with four times less particles are computationally a lot cheaper. In this subsection, we will compare these low resolution simulations to the high resolution simulations for two distinct UVB models and discuss the convergence of the results.

Note that there is almost a factor of 10 difference between the masses of the lowest mass and the highest mass ICs we employ. This means that there will also be a factor of 10 difference between the effective mass resolution of these simulations. Likewise there is a difference in the mass resolutions used for baryons and dark matter, since we use the same number of particles for both components. Since the run time of the simulation is set by the actual number of particles and not by their masses, we will use the same number of particles for ICs with different masses. We expect the resolution to be worst for the galaxies with the highest total mass.

We are interested in global properties of the final dwarf galaxy~: total stellar mass, neutral gas mass and the location of the galaxy on the BTFR. These properties are mainly set by two mechanisms. On the one hand, the star formation itself will govern the amount of neutral gas converted into stars, which affects both the stellar mass and the neutral gas mass. On the other hand, feedback from already formed stars will affect the neutral gas mass and through the gravitational interaction between gas and dark matter will affect the galaxy potential and hence the circular velocity of the halo. To obtain a converged result, we hence need sufficient resolution to (a) get the global SFH right, and (b) correctly resolve the effect of stellar feedback on the gas surrounding young star particles.

In \sectionref{subsection_stochasticity}, we concluded that, for the low resolution models, stochastic effects do not cause significant differences in final stellar mass, circular velocity or being able to keep a significant amount of neutral gas. They do however affect the final neutral gas mass if it is low compared to the stellar mass. Since ICs with a different number of particles inevitably will suffer from the same stochastic effects, we should keep this in mind when comparing the results for different resolutions.

\begin{figure}
\centering
\includegraphics[width=0.5\textwidth]{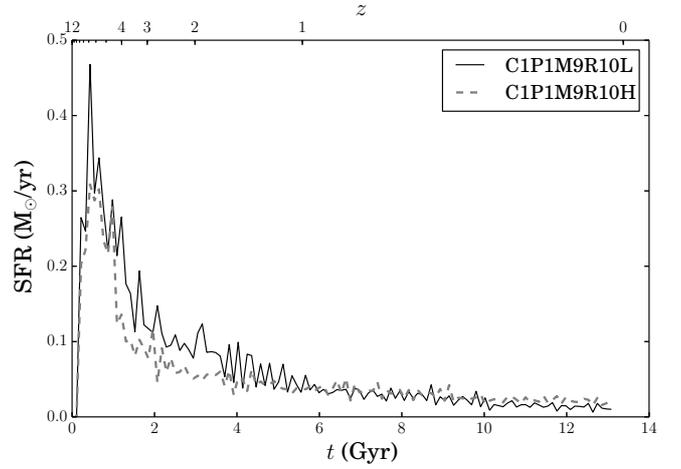}
\caption{Star formation history for the same model, but with different resolutions. The full line represents the low resolution version, the dashed line represents the high resolution version.\label{fig_convergence_SFR}}
\end{figure}

\figureref{fig_convergence_SFR} shows the typical SFH for one of our models and for both the low resolution and the high resolution version of the simulation. Qualitatively, the SFH is very similar~: there is a large peak in star formation at the start of the simulation, after which the feedback of these initially formed stars strongly suppresses further star formation by dispersing the neutral gas. Heating by the UVB prevents this dispersed gas from falling in again, resulting in a very low ongoing star formation for the remainder of the simulation. For models with a halo mass below $5\times 10^9\msol{}$, all neutral gas is removed after the initial star formation peak, so that star formation shuts down entirely. This behaviour is reproduced by both the low and high resolution runs and also depends upon the rotation parameter of the halo (see \sectionref{subsection_UVB}). It is also in agreement with the results of \citet{simpson}.

\begin{figure}
\centering
\includegraphics[width=0.5\textwidth]{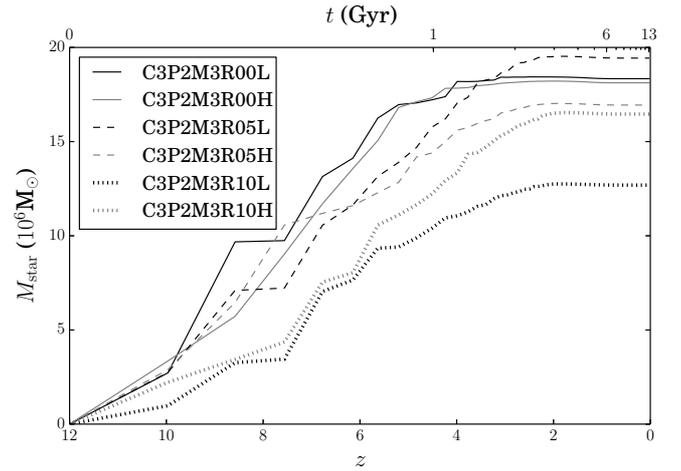}
\caption{Cumulative star formation history for the model with early and full UVB and high feedback efficiency, a halo mass of $3\times 10^9\msol{}$ and the three different rotation parameters~: no rotation (full lines), $5 \kms{}$ (dashed lines), $10 \kms{}$ (dotted lines). The black lines represent the low resolution version, the grey lines the high resolution version.\label{fig_convergence_cumSFH}}
\end{figure}

Qualitatively, there is hence no difference between the high and the low resolution runs. Both form the bulk of their stellar mass during the first gigayears of the simulations, as can be seen from \figureref{fig_convergence_cumSFH}. The simulations with a higher rotation parameter build up their stellar mass somewhat slower and this behaviour is reproduced for both the low and the high resolution versions.

\begin{figure}
\centering
\includegraphics[width=0.5\textwidth]{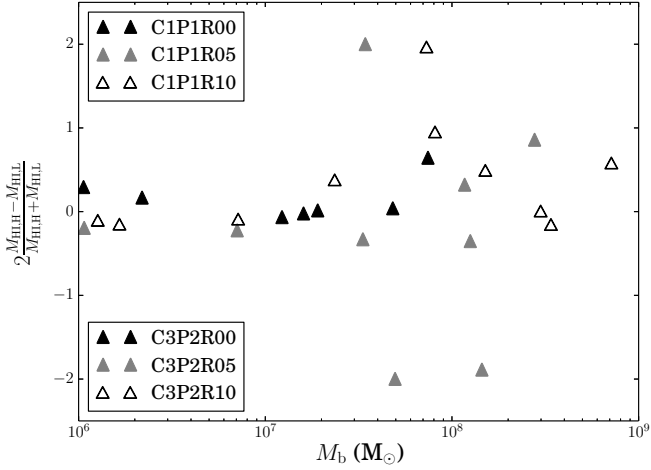}
\caption{The relative difference of the final neutral gas mass as a function of the average final baryonic mass for the simulations using the same model, but with different resolutions. The symbols correspond to the different model and initial condition parameters as indicated in the legend.\label{fig_convergence_gtfr}}
\end{figure}

\begin{figure}
\centering
\includegraphics[width=0.5\textwidth]{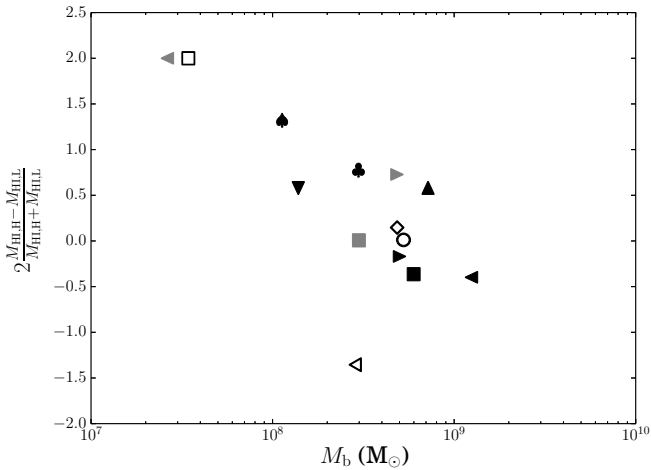}
\caption{The relative difference of the final neutral gas mass as a function of the average final baryonic mass for all models (except model CeP1), for every model comparing simulation M9R10L with simulation M9R10H. The symbols represent the different code and parameter values following \tableref{table_codes}.\label{fig_convergence_all}}
\end{figure}

\figureref{fig_convergence_gtfr} shows the relative difference in final neutral gas mass for the two models between both resolutions. We notice the same pattern as in the case of stochastic differences between the initial conditions~: the relative difference is small in the low and high mass regimes, but is significant in the intermediate mass regime. Again, this is due to the stochastic nature of feedback in the simulations. If the overall neutral gas mass is low, not much feedback is needed to disperse this gas, so small stochastic differences can lead to significantly different neutral gas masses.

As a result, models with relatively more feedback and less neutral gas will have larger differences between the low resolution and high resolution version, as illustrated in \figureref{fig_convergence_all}. Model CeP1 is excluded from this figure, as simulation CeP1M9R10H exceeded the 3 month time limit. However, both CeP1M9R10L and CeP1M9R10H have no neutral gas at the latest time a snapshot was written for CeP1M9R10H.

\begin{figure}
\centering
\includegraphics[width=0.5\textwidth]{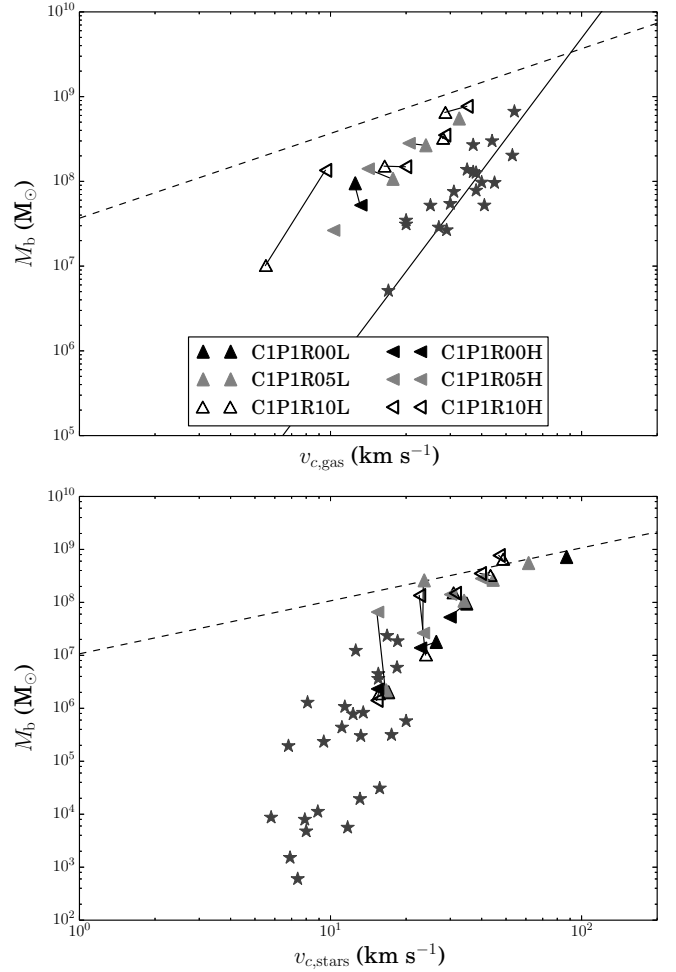}
\caption{Baryonic Tully-Fisher relation for the models with late and low UV background and low stellar feedback efficiency. The upward facing triangles correspond to the low resolution versions, the left facing triangles correspond to the high resolution versions. The stars and the full line represent the observational data and a fit to it from \citet{mcgaugh}, the other symbols are our models, as indicated in the legend. The dashed lines correspond to the least-squares fit to all simulations in \figureref{fig_btfr_all}. For clarity, simulations representing the same initial condition have been joined by a line.\label{fig_convergence_BTFR_C1P1}}
\end{figure}

\begin{figure}
\centering
\includegraphics[width=0.5\textwidth]{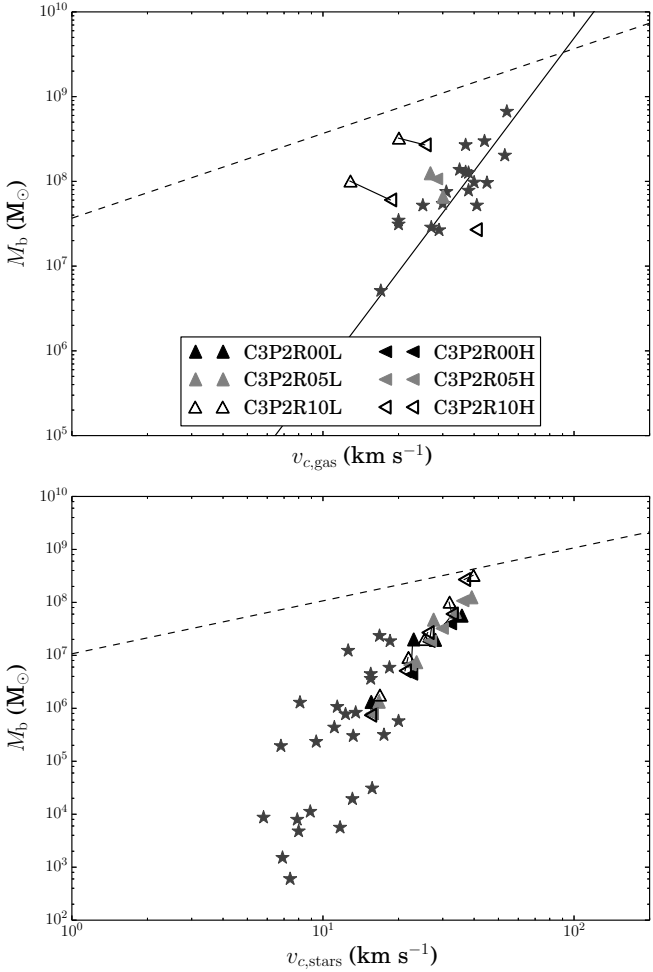}
\caption{Baryonic Tully-Fisher relation for the models with early and full UV background and high stellar feedback efficiency. The upward facing triangles correspond to the low resolution versions, the left facing triangles correspond to the high resolution versions. The stars and the full line represent the observational data and a fit to it from \citet{mcgaugh}, the other symbols are our models, as indicated in the legend. The dashed lines correspond to the least-squares fit to all simulations in \figureref{fig_btfr_all}. For clarity, simulations representing the same initial condition have been joined by a line.\label{fig_convergence_BTFR_C3P2}}
\end{figure}

The resolution has only a minor effect on the location of the simulations on the theoretical BTFR for these models, as can be seen from the bottom panels of \figureref{fig_convergence_BTFR_C1P1} and \figureref{fig_convergence_BTFR_C3P2}. For the mock observational BTFR, the resemblance is worse, but this is mainly caused by the overall low neutral gas masses, which make it very difficult to obtain reliable circular velocity estimates. For model C1P1, the mock observational BTFR looks better and in this case high and low resolution simulations trace out the same BTFR, which is higher than the observed one.

We conclude that our low resolution simulations are well converged in terms of stellar mass and theoretical circular velocity. If enough neutral gas is present, the neutral gas mass is also converged and circular velocities derived from the neutral gas are reliable. If the neutral gas mass is low, convergence is less clear, but this is in line with the effect of stochasticity. The low resolution runs are hence sufficient to distinguish between galaxies that are unable to keep any neutral gas and those that keep a significant amount of neutral gas. We can hence study the effect of different parameters on the BTFR using the computationally cheaper low resolution simulations and expect our results to be qualitatively correct. For more detailed studies of other properties of the dwarf galaxies, especially related to the small scale structure of the gas halo, we should use the more expensive high resolution runs, but this falls outside the scope of this work.

\begin{figure}
\centering
\includegraphics[width=0.5\textwidth]{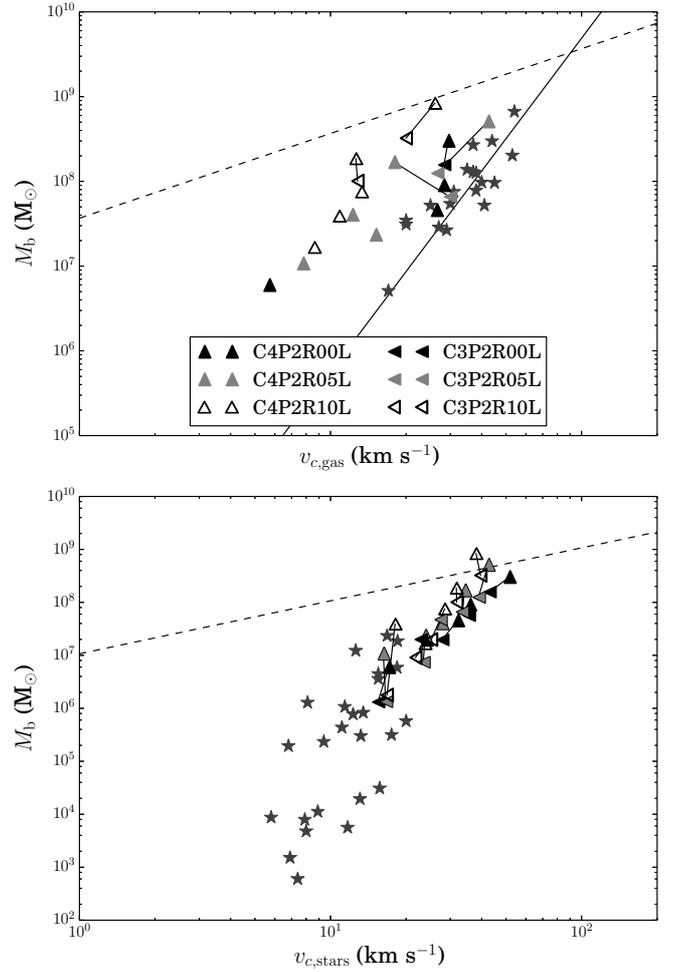}
\caption{The baryonic Tully-Fisher relation for the models without a UV background and the models with a full and early UV background with the same stellar feedback strength. The stars and the full line represent the observational data and a fit to it from \citet{mcgaugh}, the other symbols are our models, as indicated in the legend. The dashed lines correspond to the least-squares fit to all simulations in \figureref{fig_btfr_all}. For clarity, simulations representing the same initial condition have been joined by a line.\label{fig_UVB_BTFR_C4P2}}
\end{figure}

\subsection{UV background}\label{subsection_UVB}
Without a UVB, all our models form too many stars and are able to keep a significant amount of neutral gas, making them trace out a BTFR that consistently lies above the observed relation \figurerefp{fig_UVB_BTFR_C4P2}. We need a high stellar feedback parameter (1.0) to somewhat suppress star formation in these models and even this high value is clearly not sufficient \citep{annelies_sfp}.

\begin{figure}
\centering
\includegraphics[width=0.5\textwidth]{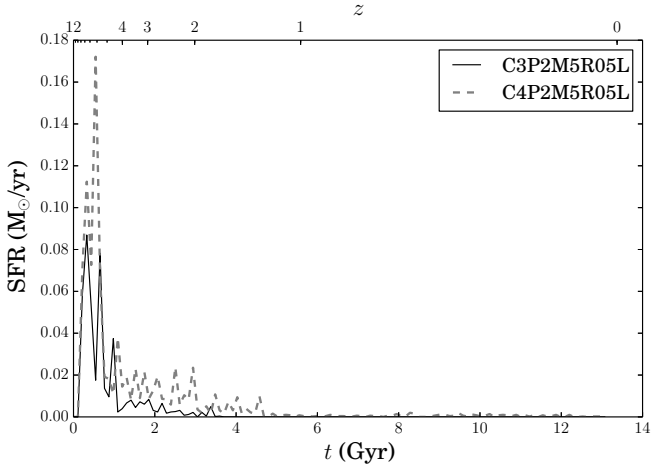}
\caption{The star formation history for a model with (full line) and without (dashed line) UVB.\label{fig_UVB_SFH}}
\end{figure}

We already saw in \sectionref{subsection_convergence} that this behaviour dramatically changes when including a UVB~: UVB heating after the initial star formation peak quickly disperses most of the neutral gas that is heated by stellar feedback and prevents the gas from cooling and falling in again. This means that the SFH is qualitatively different from the one in the reference model without UVB \figurerefp{fig_UVB_SFH}. After the initial peak, further star formation is strongly suppressed in the models with UVB and even completely stops in the lowest mass models.

\begin{figure}
\centering
\includegraphics[width=0.5\textwidth]{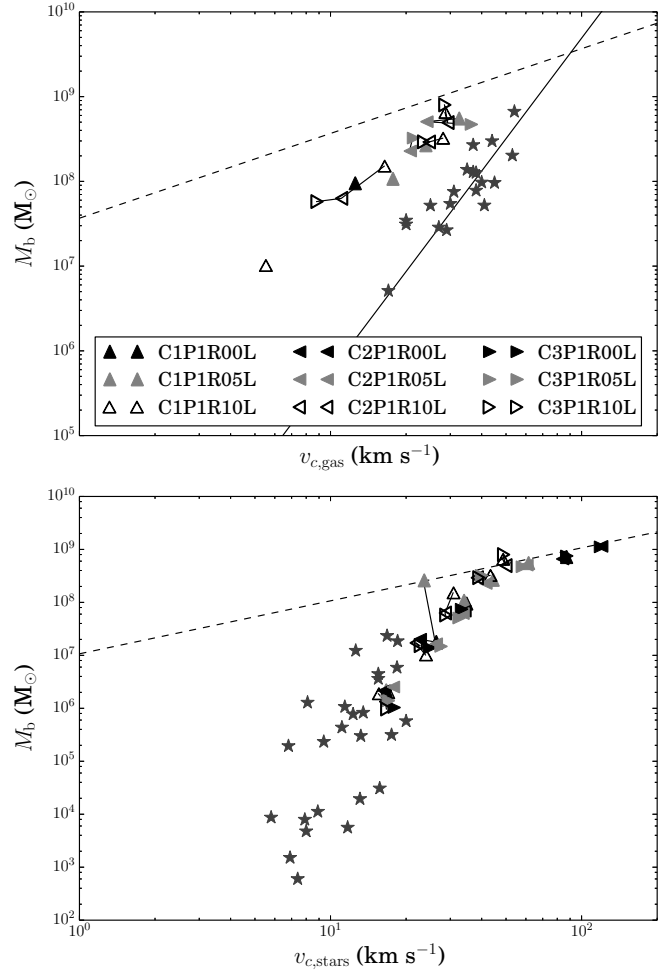}
\caption{The baryonic Tully-Fisher relation for the three different UVB models. The upward facing triangles correspond to the model with a late and low UVB, the left facing triangles correspond to the model with a late and full UVB, and the right facing triangles correspond to the model with an early and full UVB. The stars and the full line represent the observational data and a fit to it from \citet{mcgaugh}, the other symbols are our models, as indicated in the legend. The dashed lines correspond to the least-squares fit to all simulations in \figureref{fig_btfr_all}. For clarity, simulations representing the same initial condition have been joined by a line.\label{fig_UVB_BTFRs}}
\end{figure}

When comparing different models for the UVB, the BTFR turns out to be very resilient against changes in UVB model. This is illustrated in \figureref{fig_UVB_BTFRs}, where we compare models with respectively a late and low UVB, a late and full UVB, and an early and full UVB (see \sectionref{subsection_UVB_model} for the parameters of these models). The three models trace out a very similar BTFR.

The reason for this resilience is twofold. On the one hand, the strength of the UVB heating does not scale linearly with the strength of the UVB. For instance, when the UVB intensity is decreased, the neutral fraction of the gas will go up, which means there is more neutral gas that can absorb UV radiation. As a result, the UVB heating rate will be almost unchanged. If we really want to lower the strength of the heating with a considerable factor, we would need to use unphysically low UVB intensities. On the other hand, the timing of the onset of the UVB is completely masked by the first star formation peak. Although there is a significant difference in redshift between an early and a late UVB, there is only a small difference in time between these events at these high redshifts. This means that the UVB will still kick in early in the simulation, while the gas is collapsing and causing the first large star formation peak. The effect of either UVB on the galaxy will only become noticeable after this first peak, when the gas is dispersed by feedback from the first stars. At this moment, both the early and the late UVB will already have kicked in and their strengths will be comparable.

\subsection{Merger simulation}

\begin{figure}
\centering{}
\includegraphics[width=0.5\textwidth]{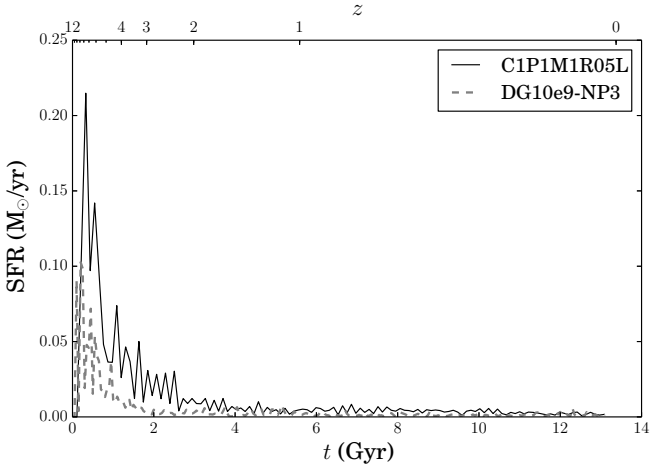}
\caption{The star formation history for an isolated model (full line) and for the merger simulation (dashed line). Both models have a comparable final halo mass of $\sim{}10^9\msol{}$.\label{fig_merger_sfh}}
\end{figure}

It is clear that many of the problems with our models are caused by the large star formation peak at the beginning of the simulation. To check whether this initial peak is caused by the idealized initial conditions that neglect cosmological effects, we will investigate the star formation history of the merger simulation DG10e9-NP3. This is shown in \figureref{fig_merger_sfh}. From \figureref{fig_btfr_all}, it is already clear that this model lies significantly above the observational BTFR.

Compared to an isolated model with a comparable final halo mass, the merger simulation shows an overall lower star formation rate, so that the final stellar mass is roughly a factor two lower than for the isolated model. The initial star formation peak however still stands out as a clear feature in the star formation history.

We conclude that cosmological effects alone are not enough to reduce the initial star formation peak. We do however expect that models that are able to reduce this peak in isolated simulations, will perform even better when cosmological effects are taken into account \citep{verbeke2015}. This falls outside the scope of this work, where we focus on the effect of internal feedback mechanisms on the initial star formation peak.

\subsection{Over-cooling}

\begin{figure}
\centering
\includegraphics[width=0.5\textwidth]{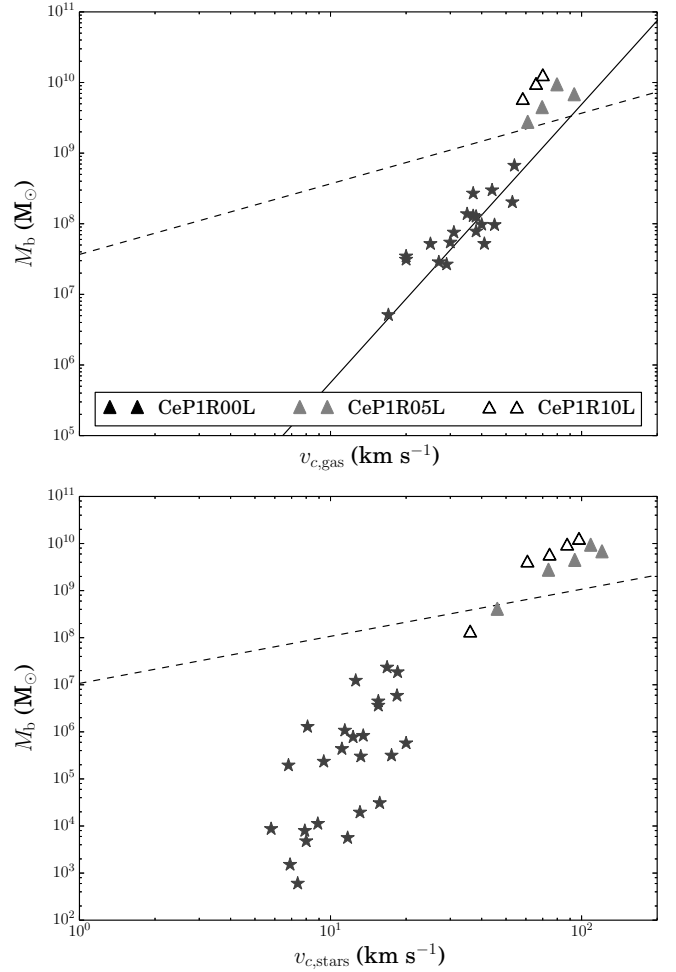}
\caption{The baryonic Tully-Fisher relation for the simulations not including an adiabatic cooling period for gas receiving stellar feedback. The stars and the full line represent the observational data and a fit to it from \citet{mcgaugh}, the other symbols are our models, as indicated in the legend. The dashed lines correspond to the least-squares fit to all simulations in \figureref{fig_btfr_all}.\label{fig_overcooling_BTFR}}
\end{figure}

Before we can investigate the effect of varying the stellar feedback efficiency parameter, we have to address the effect of over-cooling on our simulations. As discussed in \sectionref{subsection_model_stellar_feedback}, we ad hoc switch off cooling for gas particles that receive feedback from SW and SNII, to allow them to go through a phase of adiabatic cooling before radiative cooling starts to radiate away the feedback energy. If we would not do this, most of the feedback energy would be radiated away immediately, leading to very inefficient feedback. The resulting galaxies form way too many stars and are far above the observed BTFR, as shown in \figureref{fig_overcooling_BTFR}.

\begin{figure}
\centering{}
\includegraphics[width=0.5\textwidth]{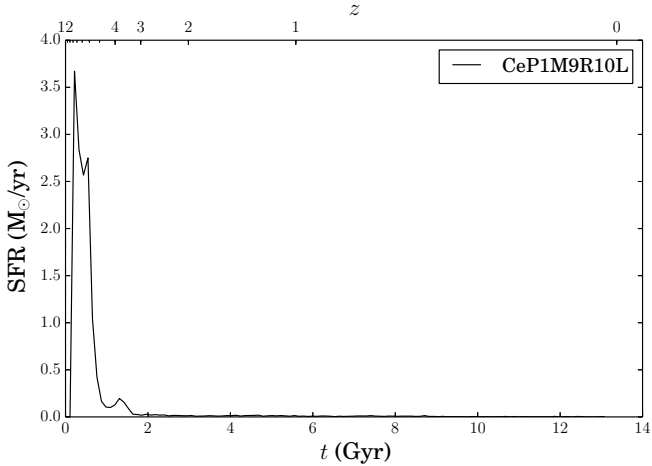}
\caption{The star formation history for a model without an adiabatic cooling period for gas that receives stellar feedback.\label{fig_overcooling_SFR}}
\end{figure}

The excessive star formation is entirely due to the first peak in star formation, which immediately consumes all the gas in the simulation, as illustrated by the SFH in \figureref{fig_overcooling_SFR}.

\subsection{Stellar feedback efficiency}\label{subsection_feedback}
In \sectionref{subsection_UVB}, we discussed the influence of an external heating mechanism on the BTFR, and we showed that, if this external heating is provided by an UVB, this influence is very small. In this part, we will examine the influence of internal heating processes, which are due to feedback from massive stars.

For the bulk of our simulations, we used a feedback parameter of 0.7. To assess the influence of the parameter, we compare it with a model with respectively a feedback parameter of 1.0 and of 2.0. The latter is strictly speaking unphysical, although due to numerical resolution issues we cannot guarantee that all energy that is put into the gas is ultimately used to heat the gas, so in this sense a value of 2.0 might be justified.

\begin{figure}
\centering
\includegraphics[width=0.5\textwidth]{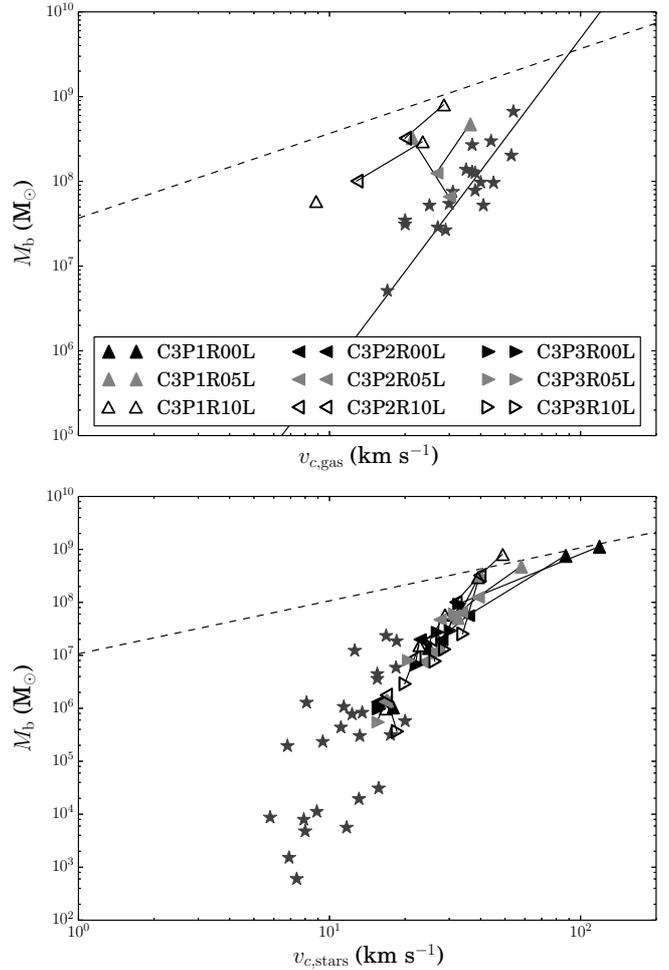}
\caption{The baryonic Tully-Fisher relation for the three different feedback parameter values. The upward facing triangles correspond to the model with feedback parameter 0.7, the left facing triangles correspond to the model with feedback parameter 1.0, and the right facing triangles correspond to the model with feedback parameter 2.0. The stars and the full line represent the observational data and a fit to it from \citet{mcgaugh}, the other symbols are our models, as indicated in the legend. The dashed lines correspond to the least-squares fit to all simulations in \figureref{fig_btfr_all}. For clarity, simulations representing the same initial condition have been joined by a line.\label{fig_feedback_BTFRs}}
\end{figure}

The results of the comparison are shown in \figureref{fig_feedback_BTFRs}. The effect is considerable~: the models with a higher feedback parameter generally have a lower baryonic mass and the more massive models tend to also have lower circular velocities. The models with feedback parameter 2.0 are completely absent on the mock observational BTFR, since none of them has any neutral gas mass left at the end of the simulation.

\begin{figure}
\centering
\includegraphics[width=0.5\textwidth]{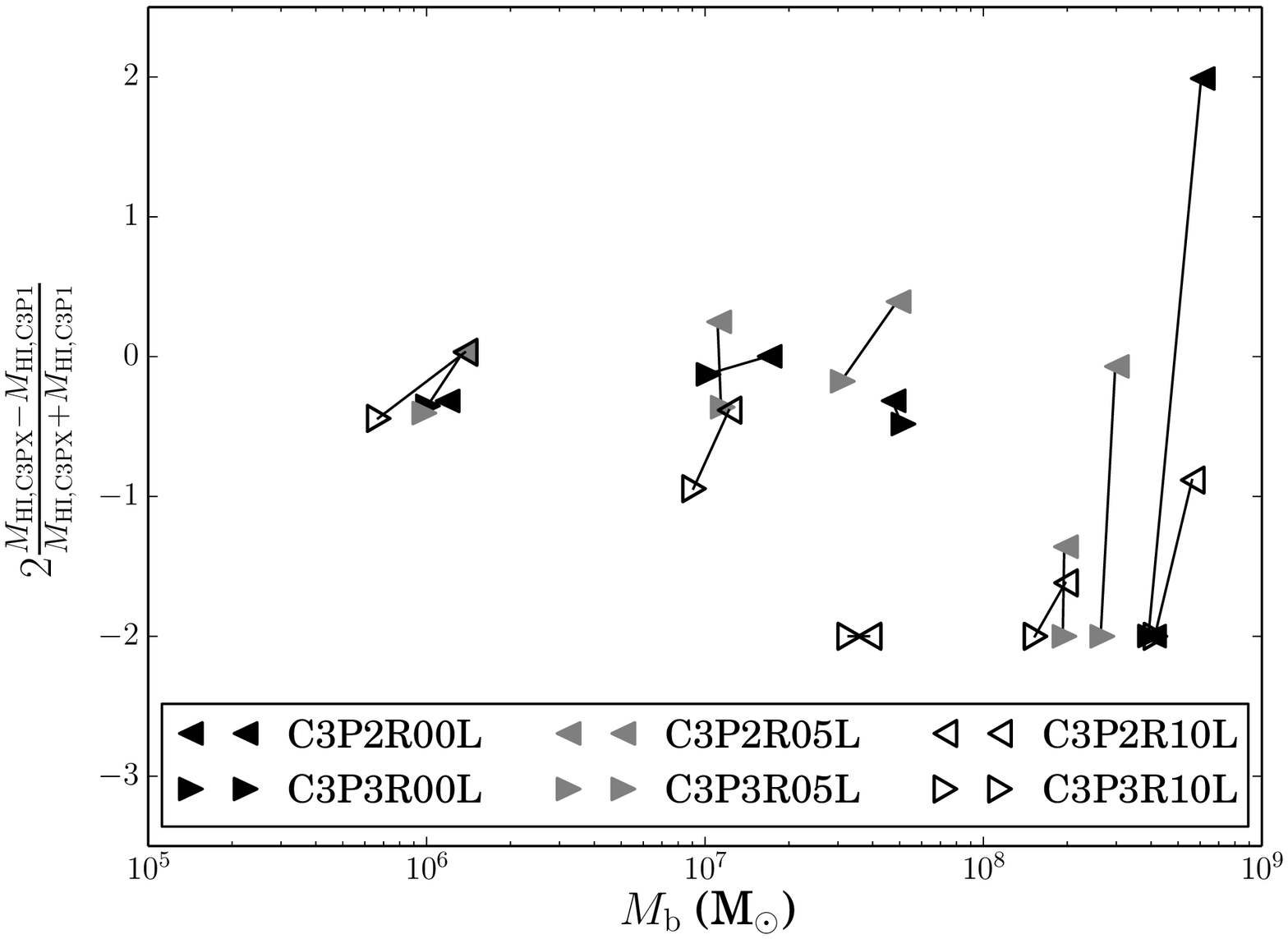}
\caption{The relative difference of the final neutral gas mass as a function of the average final baryonic mass for the simulations with different feedback parameters. The left facing triangles correspond to the relative difference between the models with feedback parameter 1.0 and 0.7, the right facing triangles to the relative difference between the models with feedback parameter 2.0 and 0.7. For clarity, simulations representing the same initial condition have been joined by a line.\label{fig_feedback_GTFRs}}
\end{figure}

If we only look at the final neutral gas mass of the models \figurerefp{fig_feedback_GTFRs}, we see that the feedback parameter regulates the cut-off mass for which halos are no longer able to keep neutral gas~: this cut-off mass is higher for the models with a higher feedback parameter.

Decreasing the feedback parameter below 0.7 will only shift the BTFR to higher baryonic masses, while it is already too high at the high velocity end. Although we could expect this to lead to more realistic final neutral gas masses, it will also lead to excessive stellar masses.

Apart from slightly shifting the total stellar mass and the circular velocity to lower values, increasing the feedback parameter does hence not help to produce galaxies that lie on the observational BTFR and even has a detrimental effect on the final neutral gas mass.

\subsection{\popthree{} feedback}
In \sectionref{subsection_UVB} and \sectionref{subsection_feedback}, we discussed the effect of external and internal heating on the resulting BTFR for our models. For the case of the external UVB, we considered two aspects of the external feedback~: its strength and its timing. Both are well constrained and hence leave us little to play with. For the internal stellar feedback, we only discussed the strength. The timing of the feedback is in this case also well constrained~: stellar winds and SNII feedback occur shortly after the star population was born, while SNIa feedback occurs with some delay and is more spread out over time. There is no reason why the feedback strength should vary with time, since temporal changes in energy absorption in the ISM are handled by the metallicity dependence of our gas cooling model.

An exception to this are \popthree{} stars, which are born from very low metallicity gas early in the simulation and have a different feedback compared to other stars. Since they enrich the ISM with their feedback, \popthree{} stars will only be born early on in the simulation, which effectively causes the stellar feedback to change over time. This could provide a solution to the problems caused by the initial large star formation peak, since this gives us a mechanism to reduce this first peak without affecting star formation in later stages of the simulation.

\begin{figure}
\centering{}
\includegraphics[width=0.5\textwidth]{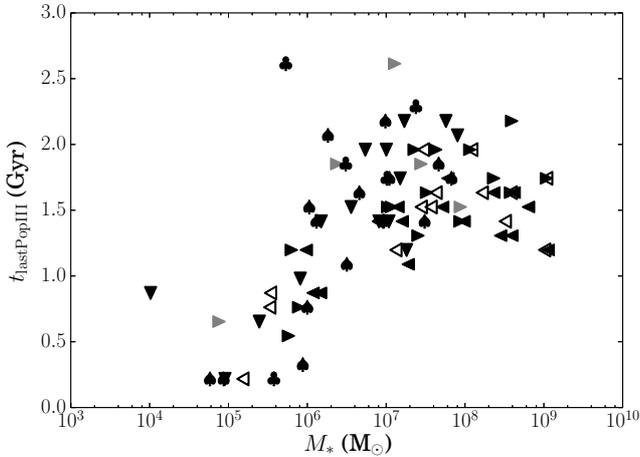}
\caption{The latest formation time of a \popthree{} star as a function of the final stellar mass for all simulations including \popthree{} feedback. The symbols represent the different code and parameter values following \tableref{table_codes}.\label{fig_PopIII_stats}}
\end{figure}

\figureref{fig_PopIII_stats} shows the latest formation time of a \popthree{} star as a function of the final stellar mass for all simulations including \popthree{} feedback. It is clear that \popthree{} stars are indeed only formed at the early stages of the simulations, so that \popthree{} feedback is limited to these early stages.

In \sectionref{subsection_popthree_models} we introduced three models for \popthree{} feedback. Model 1 only affects the SN feedback from \popthree{} stellar particles, model 2 also includes a \popthree{} SW, while model 3 implements more advanced metal yields. We will discuss the effect of the different models separately.

\subsubsection{Model 1}
Model 1 has two variants~: variant A assumes a lower limit of 60$\msol{}$ on the mass of \popthree{} stars, while variant B has a lower limit of 140$\msol{}$ and hence returns its energy over a much shorter time interval.

\begin{figure}
\centering
\includegraphics[width=0.5\textwidth]{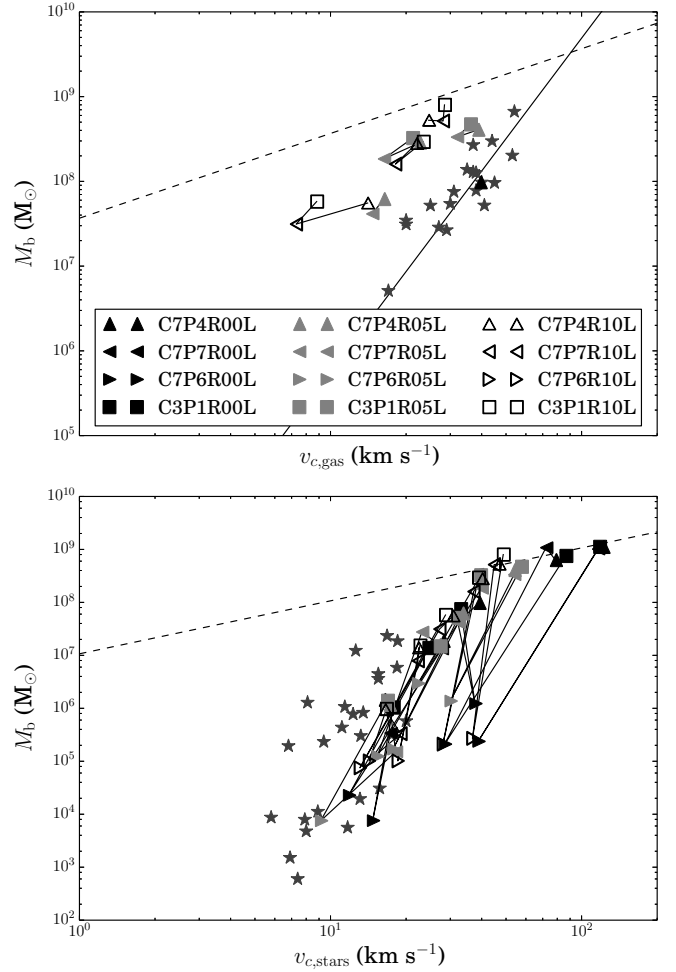}
\caption{The baryonic Tully-Fisher relation for the three different feedback parameter values for \popthree{} model 1A. The upward facing triangles correspond to the model with low feedback, the left facing triangles correspond to the model with middle feedback, and the right facing triangles correspond to the model with high feedback. The squares correspond to the model with the same parameters and no \popthree{} feedback. The stars and the full line represent the observational data and a fit to it from \citet{mcgaugh}, the other symbols are our models, as indicated in the legend. The dashed lines correspond to the least-squares fit to all simulations in \figureref{fig_btfr_all}. For clarity, simulations representing the same initial condition have been joined by a line.\label{fig_popIII_model1a_BTFRs}}
\end{figure}

\begin{figure}
\centering
\includegraphics[width=0.5\textwidth]{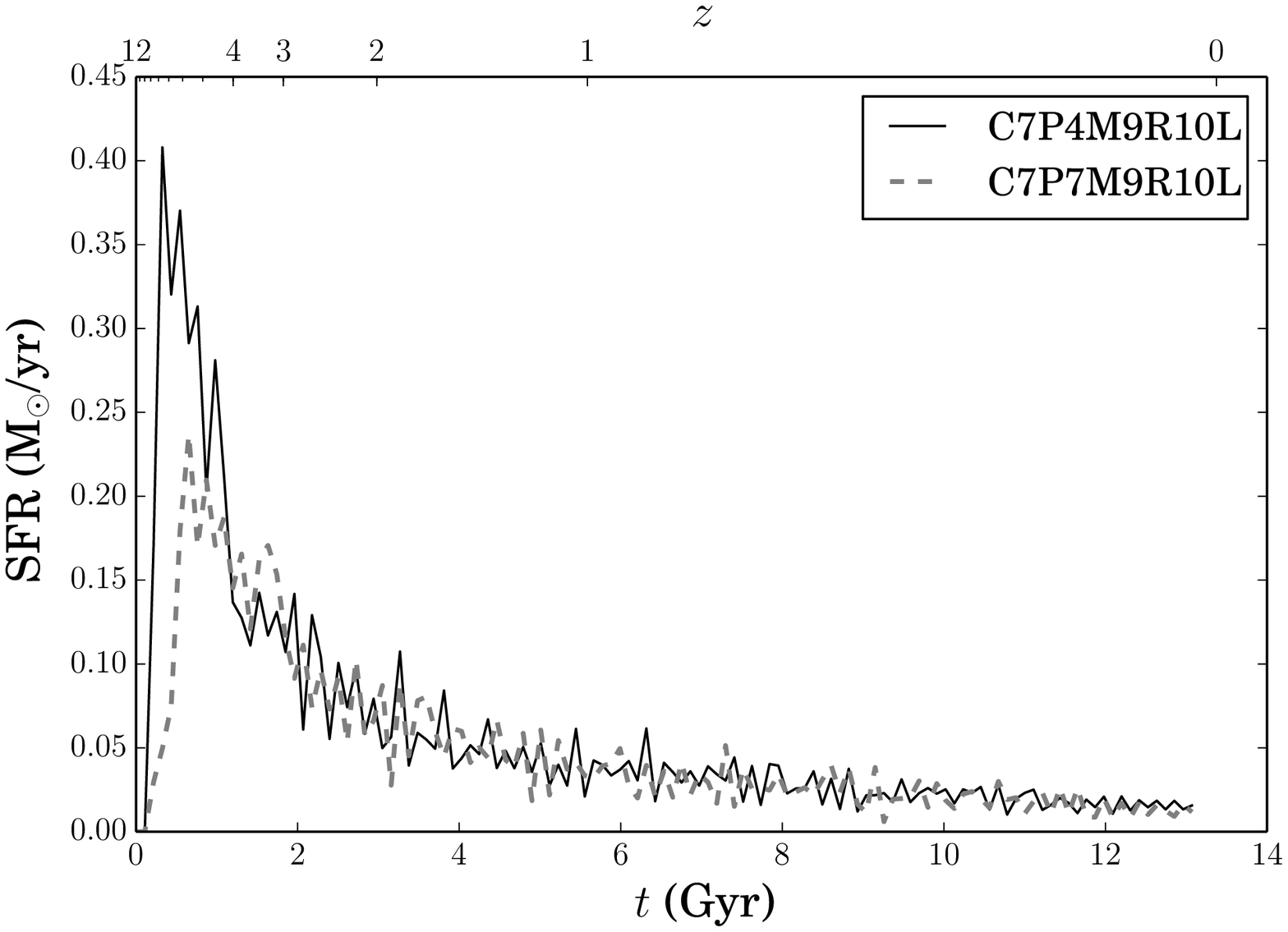}
\caption{The SFR for \popthree{} model 1A. The full line represents the model with low \popthree{} feedback, the dashed line the model with middle \popthree{} feedback.\label{fig_popIII_model1a_SFR}}
\end{figure}

\figureref{fig_popIII_model1a_BTFRs} shows the BTFR for the three realizations of model 1A, which have different \popthree{} feedback energies. It is clear that the realization with a high feedback energy results in gas poor galaxies which have very little stars. The other two realization lead to similar BTFRs.

In \figureref{fig_popIII_model1a_SFR} we show the SFH for the same ICs and the low energy and middle energy realization of the model. The \popthree{} feedback effectively reduces the height of the initial star formation peak, while at the same time leaving the further star formation history untouched, apart from stochastic differences. We also notice that the initial star formation peak is not completely reduced. Further increasing the \popthree{} feedback does not work to completely reduce the peak, as we can conclude from the high feedback model.

\begin{figure}
\centering
\includegraphics[width=0.5\textwidth]{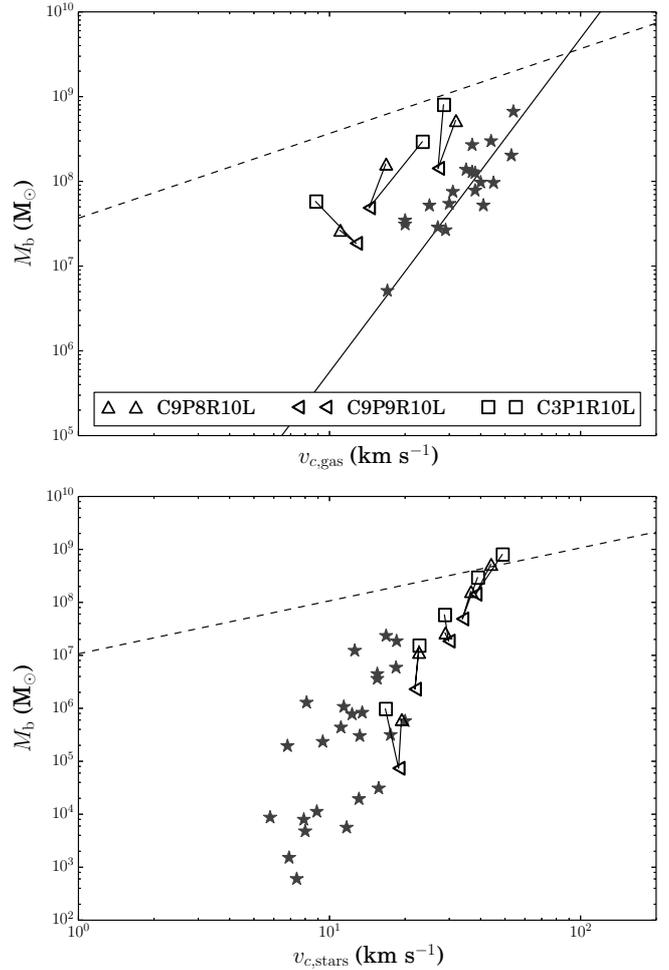}
\caption{The baryonic Tully-Fisher relation for the two different feedback parameter values for \popthree{} model 1B. The upward facing triangles correspond to the model with low feedback, the left facing triangles correspond to the model with high feedback. The squares correspond to the model with the same parameters and no \popthree{} feedback. The stars and the full line represent the observational data and a fit to it from \citet{mcgaugh}, the other symbols are our models, as indicated in the legend. The dashed lines correspond to the least-squares fit to all simulations in \figureref{fig_btfr_all}. For clarity, simulations representing the same initial condition have been joined by a line.\label{fig_popIII_model1b_BTFRs}}
\end{figure}

\begin{figure}
\centering
\includegraphics[width=0.5\textwidth]{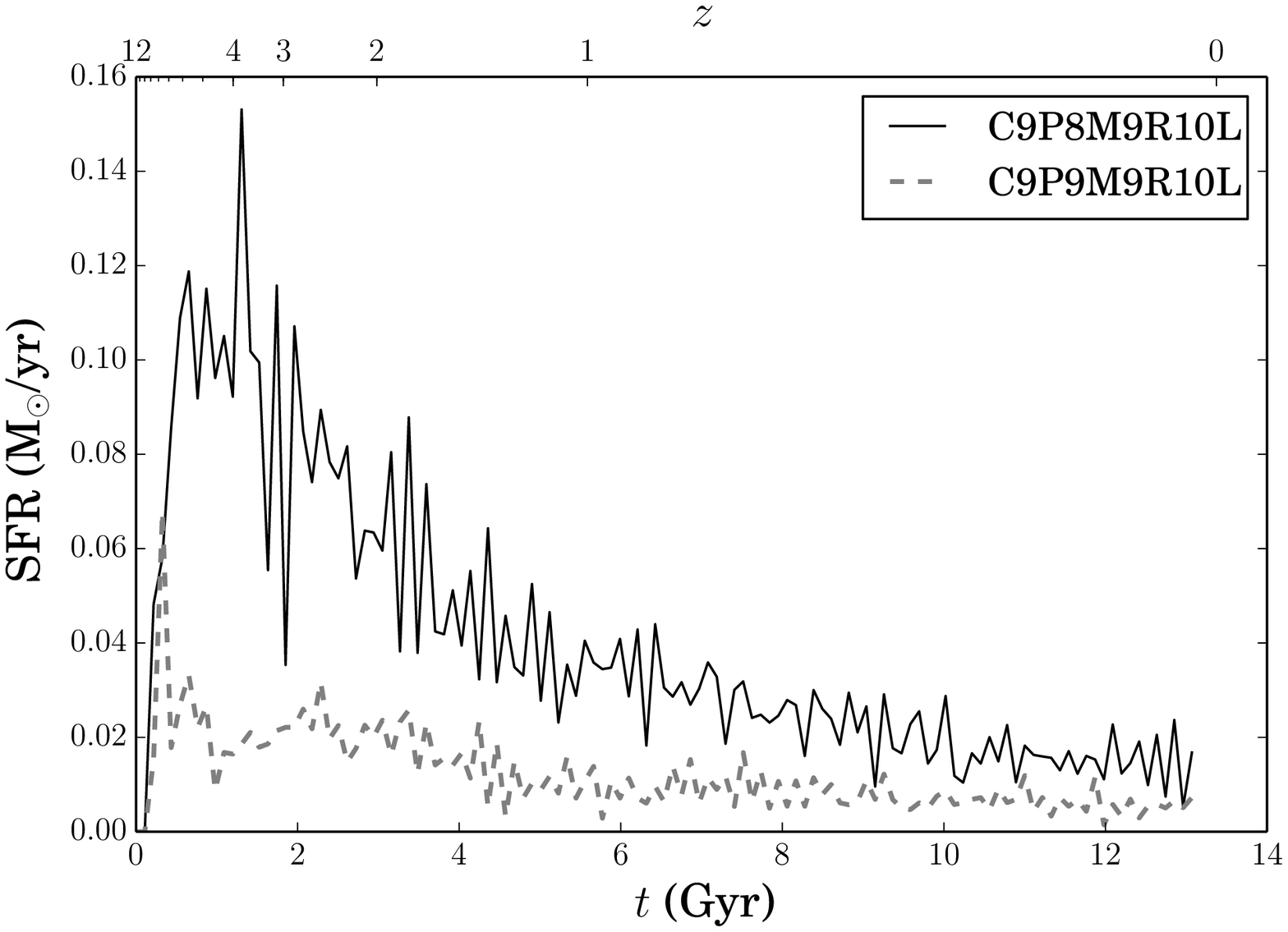}
\caption{The SFR for \popthree{} model 1B. The full line represents the model with low \popthree{} feedback, the dashed line the model with high \popthree{} feedback.\label{fig_popIII_model1b_SFR}}
\end{figure}

\figureref{fig_popIII_model1b_BTFRs} and \figureref{fig_popIII_model1b_SFR} show the BTFR and SFR for model 1B, which gives \popthree{} feedback over a much shorter time interval (and hence gives stronger feedback over a shorter period of time compared to model 1A). We see that this model effectively completely reduces the initial star formation peak.

We can conclude that the time dependence of stellar feedback introduced by \popthree{} feedback significantly changes the strength of the initial star formation peak.

\subsubsection{Model 2}

\begin{figure}
\centering
\includegraphics[width=0.5\textwidth]{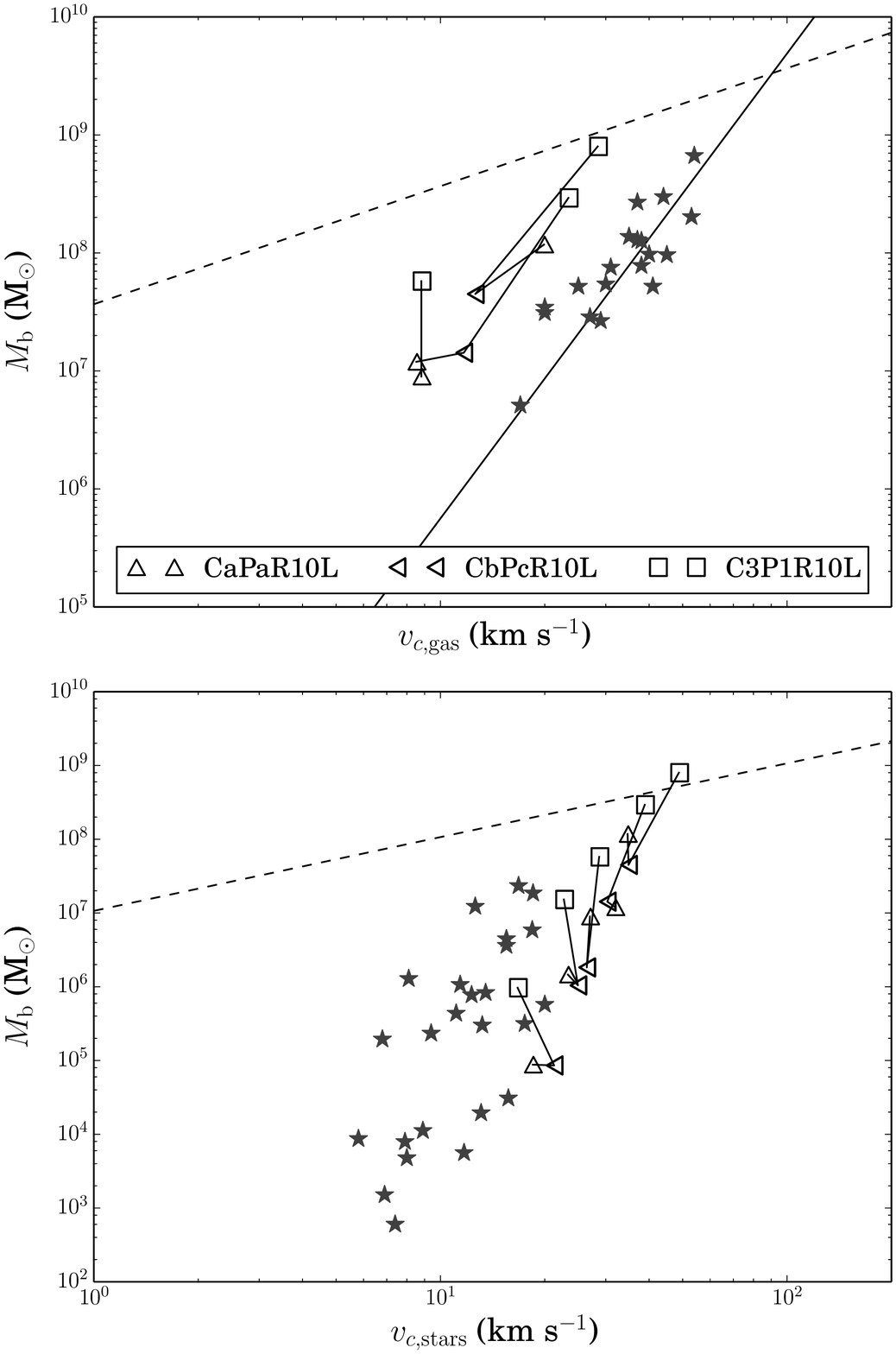}
\caption{The baryonic Tully-Fisher relation for the two different feedback parameter values for \popthree{} model 2. The upward facing triangles correspond to the model with low SN feedback and high SW feedback, the left facing triangles correspond to the model with high SN feedback and low SW feedback. The squares correspond to the model with the same parameters but no \popthree{} feedback. The stars and the full line represent the observational data and a fit to it from \citet{mcgaugh}, the other symbols are our models, as indicated in the legend. The dashed lines correspond to the least-squares fit to all simulations in \figureref{fig_btfr_all}. For clarity, simulations representing the same initial condition have been joined by a line.\label{fig_popIII_model2_BTFRs}}
\end{figure}

\begin{figure}
\centering
\includegraphics[width=0.5\textwidth]{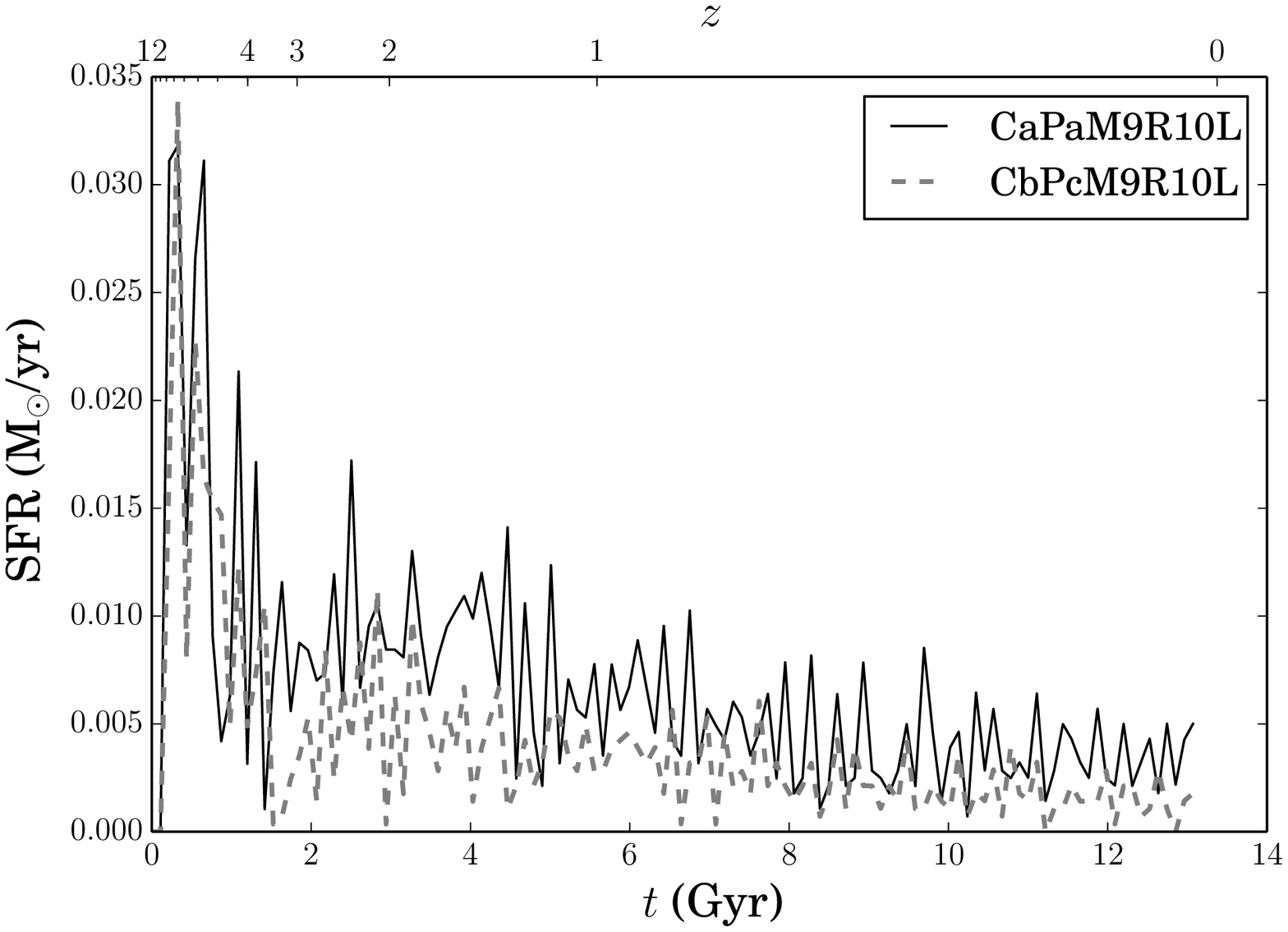}
\caption{The SFR for \popthree{} model 2. The full line represents the model with low \popthree{} SN feedback and high SW feedback, the dashed line the model with high \popthree{} SN feedback and low SW feedback.\label{fig_popIII_model2_SFR}}
\end{figure}

To completely suppress the initial star formation peak, we have to assume \popthree{} feedback over an extremely short time interval, which corresponds to unrealistically high masses for \popthree{} stars. We can get the same effect if we use a more realistic \popthree{} mass range and also include \popthree{} SW. \figureref{fig_popIII_model2_BTFRs} and \figureref{fig_popIII_model2_SFR} show the BTFR and SFH for the two models including \popthree{} SW. One model has high SW feedback and low SN feedback and vice versa. Both show a significant suppression of the initial star formation peak and further star formation and are able to keep realistic amounts of neutral gas for the more massive ICs. However, they both form too many stars and have very low circular velocities.

\subsubsection{Model 3}

\begin{figure}
\centering
\includegraphics[width=0.5\textwidth]{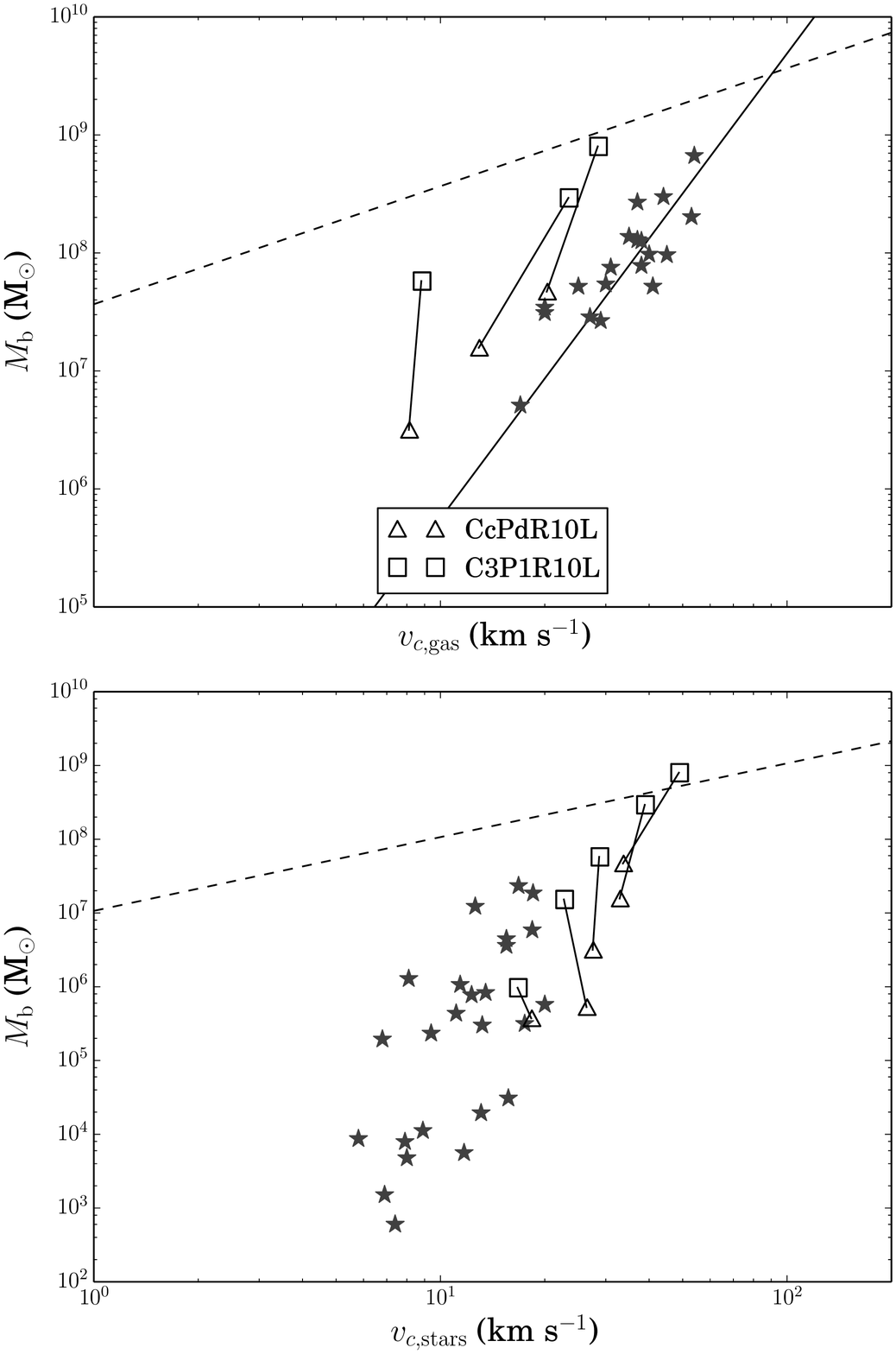}
\caption{The baryonic Tully-Fisher relation for \popthree{} model 3. The squares correspond to the model with the same parameters but no \popthree{} feedback. The stars and the full line represent the observational data and a fit to it from \citet{mcgaugh}, the other symbols are our models, as indicated in the legend. The dashed lines correspond to the least-squares fit to all simulations in \figureref{fig_btfr_all}. For clarity, simulations representing the same initial condition have been joined by a line.\label{fig_popIII_model3_BTFRs}}
\end{figure}

In the last model, we also include \popthree{} metal yields in our feedback model, taking into account the fact that \popthree{} stars will return less metals to the ISM than \popone{} and \poptwo{} stars. This in turn will affect the cooling and heating of the gas and can influence the further star formation. The results of this model are very similar to those of model 2, apart from some stochastic differences. They trace out the BTFR shown in \figureref{fig_popIII_model3_BTFRs}, which is still a bit higher than the observed relation, but has the same slope.

\section{Conclusion}\label{section_conclusion}
In this paper, we explored a large range of sub-grid models and parameters in order to produce simulated dwarf galaxies that lie on the observed baryonic Tully-Fisher relation and are gas rich and star forming at late redshifts, like the observed ultra-faint dIrrs Leo~P, Leo~T and Pisces~A. We found that the bulk of our simulations experience a large peak in star formation around the time the UVB kicks in, which then leads to an excess in stellar feedback that drives the pre-heated interstellar medium out of the galaxies and leaves them gas poor and dead. As a result, most simulations form too many stars, do not have enough neutral gas and consequently lie above the observed BTFR. They are also too metal poor compared to observed dwarf galaxies.

The star formation peak is unaffected by changing the timing or intensity of the UVB within a physically meaningful range due to the self-regulating character of the UVB heating and the fact that UVB heating only starts affecting the galaxies at low redshifts, when the UVB is already strong.

Changing the stellar feedback strength affects the absolute height of the initial star formation peak and consequently shifts the resulting galaxies on the BTFR, but does not help in reducing the relative height of the peak with respect to the consecutive star formation, mainly due to the self-regulating character of the star formation.

The only way to reduce the initial star formation peak with respect to the consecutive star formation is by including a time dependence in the stellar feedback, caused by the metallicity dependent feedback of \popthree{} stars. We explored the properties this \popthree{} feedback needs to have in order to produce realistic dwarf galaxies and found that in the most successful model, a \popthree{} stellar population returns 40 times more energy in the form of UV radiation early during its lifetime than a normal stellar population (which we call stellar wind feedback), and 4 times more energy in the form of supernova explosions from massive stars. These values are well within the current constraints from simulations of primordial stars \citep{heger_and_woosley, nomoto}. Furthermore, a \popthree{} stellar population loses 45 per cent of its mass due to stellar feedback, of which 0.009327 per cent is Fe and 0.01514 per cent is Mg.

We note that the time dependence of the stellar feedback introduced by \popthree{} feedback is similar to introducing a time dependent star formation efficiency as discussed in \citet{krumholz}, although it is unclear whether the effect of the latter will be as strong.

With our new \popthree{} feedback prescriptions, we succeed in reproducing the slope of the observed BTFR and we obtain galaxies with more realistic metallicities. Our simple model does however not completely solve the problems discussed above, because the simulations still form slightly too much stars and loose too much neutral gas. This is mainly caused by the fact that we still consider galaxies in isolation, while galaxies in a $\Lambda$CDM Universe are formed by subsequent mergers of smaller halos, which has considerable effect on the properties of these galaxies \citep{karman}. \citet{annelies_mergers} showed that including the merger history of a halo also helps reducing the initial star formation peak, leading to galaxies that have more realistic stellar metal contents. Although computationally more efficient than cosmological simulations, these merger simulation are not suited for a parameter study of the size considered for this work. Using our advanced sub-grid model including \popthree{} feedback for a smaller set of merger simulations is subject of Verbeke et al. (submitted).

\section*{Acknowledgments}
BV and SDR thank the Ghent University Special Research Fund for financial support. RV thanks the Interuniversity Attraction Poles Programme initiated by the Belgian Science Policy Office (IAP P7/08 CHARM).

We thank Volker Springel for making \gadget{} publicly available. Our sub-grid model makes use of the CHIANTI atomic database, a collaborative project involving George Mason University, the University of Michigan (USA) and the University of Cambridge (UK).

\bsp	
\label{lastpage}
\end{document}